\documentclass[11pt]{article}
\usepackage{amssymb}
\usepackage{mathtools}
\usepackage{amsmath}
\usepackage{amstext}
\usepackage{graphicx,epsfig}
\usepackage{epsfig}
\usepackage{verbatim} 	
\usepackage{caption}
\usepackage{fancybox}
\usepackage{slashed}
\usepackage{color}
\usepackage{ulem}
\usepackage{enumitem}
\usepackage{subfigure}
\usepackage{parskip}
\usepackage{dsfont}
\usepackage{tabu}
\usepackage{tikz}
\usepackage{booktabs}
\usepackage[numbers,sort&compress]{natbib}
\usepackage{cancel}

\newcommand{\Comment}[1]{{}}
\definecolor{darkblue}{rgb}{0.15,0.35,0.55}
\definecolor{reddish}{rgb}{0.65, 0.2, 0.2}
\definecolor{taiga}{cmyk}{0.68,0,0.56,0.49}
\usepackage[linktocpage=true]{hyperref}
\hypersetup{
colorlinks=true,
citecolor=darkblue,
linkcolor=reddish,
urlcolor=darkblue,
pdfauthor={},
pdftitle={},
pdfsubject={}
}

\newcommand{\rd}{{\rm d}}
\newcommand{\D}{{\rm d}}

\newcommand{\bfk}{{\bf k}}

\newcommand{\bfx}{{\bf x}}

\newcommand{\be}{\begin{eqnarray} }
\newcommand{\ee}{\end{eqnarray} }
\newcommand{\bs}{\begin{split} }
\newcommand{\es}{\end{split} }

\newcommand{\R}{\mathcal{R}}
\renewcommand{\L}{\mathcal{L}}

\newcommand{\cH}{ \mathcal{H} }
\newcommand{\cN}{ \mathcal{N} }

\newcommand{\Mpl}{M_{\mathrm{Pl}}}

\newcommand{\bp}{\phi}
\newcommand{\e}{\epsilon}

\newcommand{\nn}{ \nonumber\\}

\newcommand{\dum}[2]{d_{#1}{}^{\mu_1\ldots \mu_#2}_{\nu_1\ldots\nu_{#2}}}

\title{}
\author{}

\numberwithin{equation}{section}

\begin{document}
%
~
\vspace{1truecm}
\renewcommand{\thefootnote}{\fnsymbol{footnote}}
\begin{center}
{\huge \bf{The Effective Theory of \\[0.2cm]Shift-Symmetric Cosmologies}}
\end{center} 

\vspace{2truecm}
\thispagestyle{empty}
\centerline{\Large Bernardo Finelli\footnote{\tt b.finelli@uu.nl}, Garrett Goon\footnote{\tt g.l.goon@uu.nl}, Enrico Pajer\footnote{\tt E.Pajer@uu.nl}, and Luca Santoni\footnote{\tt l.santoni@uu.nl}}
\vspace{10mm}

\centerline{\it Institute for Theoretical Physics and Center for Extreme Matter and Emergent Phenomena,}
\centerline{\it Utrecht University, Leuvenlaan 4, 3584 CE Utrecht, The Netherlands}

\vspace{10mm}

\begin{abstract}

A shift symmetry is a ubiquitous ingredient in inflationary models, both in effective constructions and in UV-finite embeddings such as string theory. It has also been proposed to play a key role in certain Dark Energy and Dark Matter models. Despite the crucial role it plays in cosmology, the observable, model independent consequences of a shift symmetry are yet unknown. Here, assuming an exact shift symmetry, we derive these consequences for single-clock cosmologies within the framework of the Effective Field Theory of Inflation. We find an infinite set of relations among the otherwise arbitrary effective coefficients, which relate non-Gaussianity to their time dependence. For example, to leading order in derivatives, these relations reduce the infinitely many free functions in the theory to just a single one. Our Effective Theory of shift-symmetric cosmologies describes, among other systems, perfect and imperfect superfluids coupled to gravity and driven superfluids in the decoupling limit. Our results are the first step to determine observationally whether a shift symmetry is at play in the laws of nature and whether it is broken by quantum gravity effects.
\end{abstract}

\newpage

\tableofcontents

\renewcommand*{\thefootnote}{\arabic{footnote}}
\setcounter{footnote}{0}

\newpage


\section{Introduction\label{Sec:Introduction}}

Successful models of inflation need to drive accelerated expansion for a prolonged period of time, traditionally a few tens of e-foldings. This is in general problematic because of the \textit{UV-sensitivity} of the inflationary mechanism. For example, for a canonical scalar field theory, one expects, on naturalness grounds, corrections to the scalar mass of the order of the cutoff of the theory $  \Lambda $, which, by consistency, must be bracketed by the Hubble parameter and the Planck scale $  \Mpl>\Lambda>H $. Further, dimension 5 and 6 operators are expected to be generated when integrating out degrees of freedom. Both of these effects generically yield large corrections to the $  \eta $ slow-roll parameter and shorten the duration of inflation to just a few efoldings (see e.g. \cite{Baumann:2014nda} for a recent review). Things get much worse for large field models with super-Planckian displacements, $  \Delta \phi>\Mpl $, where an infinite number of higher dimensional operators induce larger and larger corrections. 

It is well-known that these problems can be addressed within the effective low energy approach invoking an approximate shift symmetry, which makes the smallness of the dangerous operators technically natural. However, to yield enough efoldings of expansion, the shift symmetry needs to be preserved by Planck scale physics. This is again problematic because of the increasing circumstantial evidence that in UV-finite theories of gravity such as string theory any global symmetry should be broken \cite{ArkaniHamed:2006dz,Kallosh:1995hi,Banks:2010zn} and, even when the symmetry is eventually gauged, it may be broken by large non-perturbative effects. In this work we avoid getting into the actively debated issue of super-Planckian shift symmetries\footnote{In many papers on the topic of the Weak Gravity Conjecture \cite{ArkaniHamed:2006dz} or the Swampland \cite{Vafa:2005ui}, it is often stated that super-Planckian field ranges are problematic, namely that $  \Delta \Phi\equiv \int \rd\Phi <\Mpl$. This is most likely \textit{not} the right formulation of the problem. For example, even for an axion with sub-Planckian decay constant, it is easy to find initial conditions for which $  \Delta \Phi >\Mpl $. Also, super-Planckian curved trajectories exist in arbitrary small volumes of multifield space, e.g. along the lines of \cite{Berg:2009tg}. We therefore prefer referring to sub-Planckian shift-symmetry breaking scale rather than field displacement.}. Instead, we propose a plausible way to \textit{observationally establish} whether a shift symmetry plays a role in cosmology. We derive the model-independent consequences of an exact shift symmetry within the Effective Field Theory of Inflation \cite{Cheung:2007st}, which beautifully captures the physics of single clock cosmologies in a unifying framework. In working with the Effective Field Theory (EFT), the nature of general \textit{classes} of cosmological models is illuminated in a single fell swoop, as opposed to gleaning phenomenological predictions a single model at a time. In a future publication \cite{InPreparation}, we will extend these results to an approximate shift symmetry. Also, shift symmetric models have been proposed to model both Dark Matter (see e.g. \cite{Berezhiani:2015bqa}) and Dark Energy \cite{Celoria:2017bbh,Celoria:2017xos}. Since our analysis is general to any FLRW spacetime, it can be used to study not just inflation but these other cosmological phases as well.

It is surprising that the consequences of a shift symmetry in the well-studied EFT of Inflation have not yet been worked out\footnote{This was pointed out in the discussion of \cite{Nicolis:2011pv}, which motivated us to study this problem.}. In applications to inflation, the standard lore claims that a shift symmetry ensures that the time dependence of all the arbitrary effective coefficients must be small. This is clearly incorrect. For example, in the exact shift symmetric case slow-roll inflation is actually impossible because some slow-roll parameter and therefore some derivative of the Hubble parameter must be large (see App. \ref{App:PofXSection}). Also, even imposing the smallness of all (Hubble) slow-roll parameters, the properties of the fluid that permeates space, such as the speed of sound, can in general have a large time dependence, as happen in many condensed matter systems. In this paper, we rectify these misconceptions and show that a shift symmetry imposes infinitely many recursive relations among effective coefficients and their time derivative (see \eqref{ShiftSymmetryRelationdeltag00Models} and \eqref{recrelKt}). We recently presented in \cite{Finelli:2017fml} complementary approach using Ward-Takahashi identities and the language of adiabatic modes (see also \cite{Mooij:2015yka,Bravo:2017wyw}).

The rest of the paper is organised as follows. In Sec.~\ref{sec:flat}, as a warm up, we discuss non-gravitational systems on Minkowski spacetime and derive a single-clock Effective Theory for non-linearly realized time translations, boost and shift symmetry. This theory describes, among other systems, perfect, imperfect and driven superfluids. In Sec.~\ref{Sec:SSandEFTofI} we move on to include gravity, we review the construction of the EFT of Inflation and discuss how it changes in the presence of a shift symmetry.  In Sec.~\ref{Sec:EFTParameterConstraints} we derive the relations among EFT parameters that are dictated by the shift symmetry and in Sec.~\ref{Sec:Phenomenology} we discuss the corresponding phenomenology. We conclude in Sec.~\ref{Sec:Discussion} with a discussion.  Additionally, we provide a list of our main results below, along with our conventions.

\paragraph{Main Results} The central findings of this paper are:\begin{itemize}
\item The shift symmetry induces an infinite tower of constraints on the EFT coefficients, tying together the behavior of the standard slow-roll parameters to quantities characterizing the fluid, such as its speed of sound.  These relations are valid not only for inflationary spacetimes, but for any cosmology whose dynamics are dominated by a single scalar degree of freedom. For example, at lowest order in derivatives, the shift symmetry reduces the infinitely many arbitrary functions of time to just a single one, which can be taken to be the background evolution $ H(t)  $.
\item In the EFT of Inflation, non-linearly realized time translations elegantly imply that a small speed of sound must be accompanied with large non-Gaussianity \cite{Cheung:2007st}. Analogously, here we show that a non-linearly realized shift symmetry implies that strong time dependence must be accompanied by large non-Gaussianity.
\item A shift symmetry for a fundamental scalar $\Phi(x^{\mu})$ does not generically imply an approximate, constant symmetry for the emergent Goldstone perturbation $\pi(x^{\mu})$, as is sometimes stated.  As a result, operators such as $\sim \pi(\partial_{i}\pi)^{2}$ cannot generically be ignored in shift-symmetric models and they can play a non-negligible role in the production of primordial non-Gaussianities.
\item The fundamental shift symmetry also corresponds to the existence of an adiabatic mode, derived here, and Ward identities for its correlators, studied in \cite{Finelli:2017fml}. This is the first of many examples of adiabatic modes arising from the mixing of large diffeomorphisms and internal symmetries.
\item The construction remains interesting even in the flat space limit, where the effective theory describes the dynamics of both perfect, imperfect and driven superfluids.  The latter correspond to the study of shift-symmetric theories about non-trivial vacua in which the background scalar field profile $\bar{\Phi}(t)$ evolves as a non-linear function of time and are a generalization of the ``spontaneous symmetry probing" systems studied in \cite{Nicolis:2011pv}.
\end{itemize}

\paragraph{Conventions:}  We work in mostly plus signature and use the curvature conventions $R^{\rho}{}_{\sigma\mu\nu}=\partial_{\mu}\Gamma^{\rho}_{\nu\sigma}+\ldots$ and $R_{\mu\nu}=R^{\rho}{}_{\mu\rho\nu}$. The Hubble and slow-roll parameters are defined in cosmological time by $H\equiv \frac{\dot{a}}{a}$, $\varepsilon=-\frac{1}{H}\frac{\dot{H}}{H}$ and $\eta=\frac{1}{H}\frac{\dot{\varepsilon}}{\varepsilon}$. Spatial vectors are bolded as in $ \vec{x}\equiv\bfx$. Fourier conventions: $f(t,\bfk)\equiv \int\rd^{3}x\, e^{-i\bfk\cdot\bfx}f(t,\bfk)$, $f(t,\bfx)= \int\rd^{3}\tilde{k}\, e^{i\bfk\cdot\bfx}f(t,\bfk)$, $\tilde{k}\equiv k/(2\pi)$ and $\tilde{\delta}^{3}(\bfk)\equiv (2\pi)^{3}\delta^{3}(\bfk)$.

 
\section{Non-Gravitational, Shift-Symmetric Systems}\label{sec:flat}

In this section, we consider the introductory example of Poincar\'e invariant, shift-symmetric, non-gravitational systems and derive an Effective Field Theory for perturbations around backgrounds that non-linearly realize time translations, boosts and a shift symmetry. We also formulate these results in the language of superfluids and spontaneous symmetry probing \cite{Nicolis:2011pv}. The result can be considered as the decoupling limit, $  \Mpl\rightarrow \infty $, around Minkowski spacetime $  \bar g_{\mu\nu}=\eta_{\mu\nu} $ of the more general theory we will derive in the next sections for gravitational systems. We start with this warm up case because the absence of gauge ambiguities (diffeomorphism) makes the treatment particularly transparent. Our final EFT describes perfect, imperfect and driven superfluids.

 
\subsection{Spontaneously Broken Time Translations}\label{ssec:}

Let us start by considering a Poincar\'e invariant, non-gravitational system that non-linearly realizes time translations and boosts. We will restrict to systems with a single scalar degree of freedom.

The starting point is a scalar field $  U(x) $ that \textit{linearly} realizes Poincar\'e invariance, $  U(x)\rightarrow U(\Lambda x+\alpha) $ for some Lorentz transformation $  \Lambda $ and four-translation $  \alpha $ (suppressing Lorentz indices). The most general Lagrangian for $  U $ is 
\begin{equation}
\mathcal{L} = \mathcal{L}\left(U , \partial_\mu U , \partial_\mu\partial_\nu U, \ldots \right) \, ,
\end{equation}
where Lorentz indices are contracted with the flat spacetime metric $\eta^{\mu\nu}$. We now consider a homogeneous but time-dependent background solution $  \bar U(t) $. Using an appropriate field redefinition, one can always choose $ \bar U(t)=t  $  (assuming $\bar{U}$ is monotonic). We choose to parameterise perturbations by 
\be
U(x)\equiv \bar U( t+\pi(x))= t+\pi(x)\,,
\ee
where the field $ \pi$ can be thought of as the Goldstone field which non-linearly realizes time translations and boosts\footnote{Spacetime transformation are sometimes written in a mixed form where both the fields and the coordinates are transformed, e.g. $  t\rightarrow t+\alpha^{0} $ and $  \pi\rightarrow \pi-\alpha^{0} $. Here we stick instead to purely active transformations where only the fields are transformed. Although the two conventions are of course equivalent, we find our approach more convenient when treating systems that are not time translation invariant, because the Lagrangians have explicit time dependence.}
\be
\pi(x)&\rightarrow&\pi(\Lambda x+\alpha)+\alpha^{0}+\Lambda^{0}_{\mu}x^{\mu}-t \quad\quad \text{(Poincar\'e transformations)}\,.\label{ptra}
\ee

In the spirit of an effective description, one can organise the operators in the Lagrangian in a derivate and field expansion. This is particularly convenient because at low energies and up to some precision only a finite number of operators contribute. For instance, at the leading order in derivatives one finds the operators $d_n(U)(\partial_\mu U\partial^\mu U+1)^n$, containing at most one derivative per $U$. In term of $  \pi $, the action up to two derivatives per field reads
\be\label{flat}
S=\int \rd^{4}x \,\left[ \sum_{n=0}^{\infty} \frac{d_{n}(t+\pi)}{n!}\left(  -2\dot\pi+\partial_{\mu}\pi\partial^{\mu}\pi\right)^{n}
+\sum_{n=0}^{\infty} \frac{\hat d_{n}(t+\pi)}{n!}\left(  -2\dot\pi+\partial_{\mu}\pi\partial^{\mu}\pi\right)^{n} K
+\dots \right]\,, \label{flatspace}
\ee
where for later convenience we introduced the ``extrinsic curvature'' $K$ as\footnote{Notice that, since we are on a flat spacetime, $K$ starts linear in fluctuations, and therefore we did not have to subtract its background value in \eqref{flat}. Also, the two terms in $  K $ coincide up to a total derivative, which we have chosen to match the curved spacetime notation.}
\begin{equation}
K = \frac{\square U}{\sqrt{-\partial_\mu U\partial^\mu U}}+
\frac{\partial^\mu U\partial^\nu U \partial_\mu\partial_\nu U}{(-\partial_\mu U\partial^\mu U)^{3/2}} \, .
\end{equation}In terms of $  \pi $, $  K $ is given by (see also \eqref{deltaK})
\be
 K = -\partial^i\partial_i\pi +\dot \pi\partial^i\partial_i\pi	+ 2\partial^i\pi\partial_i\dot{\pi}	+ \mathcal{O}(\pi^3)\,. \label{deltaK2}
\ee
This form of the action obscures two useful facts: tadpole cancellation ($  d_{0}=-2d_{1}+ \text{const}$) implies that the action starts at quadratic order in perturbations and $  \pi $ decouples at zero momentum, or equivalently $  \pi=\text{const} $ is a solution of the equations of motion. It is possible to re-write \eqref{flat} so that both facts become manifest. The result is\footnote{The only term that is not manifestly quadratic is $  \partial_{i}\partial^{i}\pi $, which is a total derivative.}
\begin{multline}
S=\int \rd^{4}x \Bigg[ d_{1}(t+\pi) \partial_{\mu}\pi\partial^{\mu}\pi+\sum_{n=2}^{\infty} \frac{d_{n}(t+\pi)}{n!}\left(  -2\dot\pi+\partial_{\mu}\pi\partial^{\mu}\pi\right)^{n}
\\
 +\sum_{n=0}^{\infty} \frac{\hat d_{n}(t+\pi)}{n!}\left(  -2\dot\pi+\partial_{\mu}\pi\partial^{\mu}\pi\right)^{n}\left( -\partial^i\partial_i\pi +\dot \pi\partial^i\partial_i\pi	+ 2\partial^i\pi\partial_i\dot{\pi} + \dots \right)
+\ldots\Bigg]\,.\label{notad}
\end{multline}
The actions \eqref{flat} and \eqref{notad} describe the universal sector of any system that breaks time translations spontaneously.
Namely, the form of the operators is dictated only by the breaking pattern and no input from additional symmetries has been considered so far.

 
\subsection{Spontaneously Broken Shift Symmetry}\label{ssec:}

We now proceed to impose a shift symmetry. We assume that there exists some field redefinition $  \Phi(x)=\bar\Phi(U(x)) $, for some algebraic, single-variable function $  \bar \Phi $, such that the action \eqref{flat} is invariant under the internal symmetry $\Phi\rightarrow\Phi+c$, for some constant $c$. In terms of $\pi$, the shift-symmetry transformation is readily extracted from
\be
\Phi(x)=\bar \Phi(U(x))=\bar \Phi(t+\pi(x))\,,
\ee 
and it is found to be
\begin{equation}
\pi(x)\rightarrow\pi(x)+ \frac{c}{\dot{\bar{\Phi}}(t+\pi(x))}  \quad\quad \text{(shift symmetry)}\, ,
\label{shiftpiflat}
\end{equation}
at linear order in $  c $ but to all orders in $\pi$.
Imposing \textit{exact} invariance of the action \eqref{flat} under \eqref{shiftpiflat} provides infinitely many recursive relations among the coefficients:
\begin{align}
\frac{2\ddot{\bar \Phi}}{\dot{\bar \Phi}}  d_{n+1}&= \frac{2n\ddot{\bar \Phi}}{\dot{\bar \Phi}} d_{n}
-  \dot{d}_{n} \,, \label{cond1}\\
\frac{2\ddot{\bar \Phi}}{\dot{\bar \Phi}}  \hat d_{n+1}&= \frac{2n\ddot{\bar \Phi}}{\dot{\bar \Phi}}\hat  d_{n}
-  \dot{\hat  d}_{n} \,. \label{cond2}
\end{align}
We will extend these relations to curved spacetime in \eqref{recrelKt}.

These relations can be interpreted as the on-shell conservation $\partial_\mu J^\mu=0$ of the Noether current $J^\mu$ associated with the transformation \eqref{shiftpiflat}, expanded order by order in $\pi$. If the Lagrangian is exactly invariant under \eqref{shiftpiflat}, then the current $J^\mu = (\delta \mathcal{L}/\delta\partial_\mu\pi)\Delta\pi$ is conserved and the conditions \eqref{cond1}-\eqref{cond2} are satisfied. 
On the other hand, if the variation of the Lagrangian is a non-zero total derivative, $\Delta \mathcal{L}=\partial_\mu F^\mu$, then the conserved current is $J^\mu = (\delta \mathcal{L}/\delta\partial_\mu\pi)\Delta\pi-F^\mu$ and the relations \eqref{cond1}-\eqref{cond2} get corrected accordingly. The only non-trivial\footnote{All other terms that change by a total derivative can be integrated by parts so that they're strictly invariant.  E.g.~$\Phi\square\Phi$ shifts by a total derivative, but can be integrated to the form $-\left (\nabla\Phi\right )^{2}$.} case arises from a ``driving'' term, i.e. a tadpole $\mu\Phi$ in the Lagrangian, which generates the change $\Delta L=\mu c$. The only correction to the recursive relations takes place at $  n=0 $ in \eqref{cond1}:
\be
\frac{2\ddot{\bar \Phi}}{\dot{\bar \Phi}}  d_{n+1}= \frac{2n\ddot{\bar \Phi}}{\dot{\bar \Phi}} d_{n}
-  \dot{d}_{n} + \dot{\bar{\Phi}}\mu \delta_n^0 \, , \label{cond1-driven}
\ee
while \eqref{cond2} is unchanged. Clearly, this discussion can be further generalised to include higher derivative operators $(\partial\partial U)^n$ in the effective theory \eqref{flat}. We will comment on this point later on in the context of gravitational systems. 

It is also interesting to consider the Goldstone boson $  \varphi $ of shift symmetry transformations, defined by
\be
\Phi(x)=\bar\Phi(t)+\varphi(x)\,.
\ee
Under shift symmetry and time translations, it simply transforms as 
\be
\varphi(x)&\rightarrow&  \varphi(x)+c\hspace{6.3cm} \text{(shift symmetry)}\, ,\\
\varphi(x)&\rightarrow&\varphi(\Lambda x+\alpha) +\bar\Phi\left( \alpha^{0}+\Lambda^{0}_{\mu}x^{\mu} \right)-\bar\Phi(t)\quad\quad \text{(Poincar\'e transformations)}\, ,
\ee
$  \varphi $ is non-linearly related to the Goldstone of time-translations by
\be
\varphi(x)=\bar \Phi(t+\pi(x))-\bar \Phi(t)\,.\label{varphi}
\ee
It is worth noting that the shift symmetry \eqref{shiftpiflat} and Poincar\'e transformations \eqref{ptra} are distinct and independent from each other. In Table \ref{tab}, we summarize how the various transformations are non-linearly realized on $  \pi $ and $  \varphi $. For general $  \bar \Phi $ this is the end of the story, but one special case arises when $  \bar \Phi $ is a linear function of time \cite{Nicolis:2011pv}, in which case the system can be identified as a perfect superfluid to leading order in derivatives. In this case, $  \dot{\bar\Phi}=\text{const} $ and $  \varphi $ coincides with $  \pi $ (and so do their transformations). 

As we discuss next, the action \eqref{flatspace} with the shift symmetry relations \eqref{cond1}, \eqref{cond2} and \eqref{cond1-driven} describes the low-energy excitations of this perfect superfluid as well as more general systems such as imperfect and driven superfluids.


\subsection{Perfect Superfluids}\label{sec:PerfectSuperfluids}

In flat, non-dynamical spacetime, different states of matter can be classified according to the underlying pattern of spacetime and internal symmetry breaking (see \cite{Nicolis:2015sra} for a recent systematic discussion). From this point of view, a perfect superfluid\footnote{In this work we always implicitly assume that the superfluid is at zero temperature. A finite temperature superfluid can be described by an admixture of a perfect fluid and a zero-temperature fluid, but we will not discuss this case. See \cite{Nicolis:2011cs} for more details.} is defined as a system possessing a global internal abelian symmetry that is broken together with time translations in such a way that a diagonal combination of the two symmetries remains unbroken. To lowest order in derivatives, one finds that the low energy effective action is simply an arbitrary function $  P(X) $, where $  X\equiv -\partial_{\mu}\Phi\partial^{\mu}\Phi/2 $ for some order parameter $  \Phi $ (see e.g. \cite{Son:2002zn,Dubovsky:2005xd}). In the ground state, the order parameter takes the solution $  \Phi=t $, and so could be identified with the field $  U $. The equation of state of the superfluid fixes the form of the function $  P $ and the action indeed reproduces the correct relativistic hydrodynamic equations \cite{Carter:1999zw,khalatnikov1982relativistic}. Perturbations of a perfect superfluid are described by the action \eqref{flatspace} with $  \hat d_{n}=0 $ supplemented by the conditions \eqref{cond1}. Since $  \ddot\Phi=0 $, one simply finds 
\be
\dot d_{n}=0 \quad \text{(perfect superfluid)} \,.
\ee
This is not surprising and it has in fact been discussed in detail in \cite{Nicolis:2011pv} as a specific example of \textit{Spontaneous Symmetry Probing}: the system evolves linearly in time in a symmetry direction so nothing really evolves in time. One can show that perturbations can always be parameterized in such a way that time translations appear unbroken. Also, notice that since $  \bar\Phi $ is a linear function, the Goldstone of time translations $  \pi $ coincides with the Goldstone boson of shift symmetry transformations $  \varphi $.

For a perfect superfluid, the role played by the symmetry breaking pattern can be emphasized further: one can derive the action \eqref{flatspace} via a coset construction \cite{Nicolis:2013lma}. The starting point is the observation that, since $  \Phi $ is linear in time, a linear combination of the Hamiltonian with the generator $  Q $ of the abelian shift symmetry must be conserved. This allows one to define a modified Hamiltonian that is conserved and then proceed with the standard coset construction, which assumes an unbroken Poincar\'e group \cite{Ogievetsky}.


\subsection{Imperfect and Driven Superfluids}\label{sec:ImperfectSuperfluids}

In this subsection, we discuss two examples of systems that, like the perfect superfluid, possess a shift symmetry and are described by the action \eqref{flatspace}: the imperfect (or ``braided'') superfluid \cite{Pujolas:2011he,Deffayet:2010qz} and the driven superfluid. These systems behave very differently from the superfluid because the underlying order parameter evolves non-linearly in time, $ \bar \Phi\neq t $. As a consequence no linear combination of the time translation and shift-symmetry generators remains unbroken: the system truly evolves in time. 

The most general shift-symmetric, Poincar\'e invariant scalar theory to leading order in derivatives is a \textit{driven superfluid}
\be
\L=P(\partial_{\mu}\Phi\partial^{\mu}\Phi)-\mu \Phi\,,
\ee
for some real constant $  \mu $. The tadpole term $  \mu \Phi $ captures the coupling to some external source of shift-symmetry current, which pumps superfluid charge into the system. The equations of motion for homogeneous backgrounds are simply
\be\label{Pxeom}
\ddot\Phi\left(  P_{,X}+2XP_{,XX}\right)+\mu=0\,.
\ee
For the perfect superfluid $ \mu=0 $ and one finds\footnote{Another, less general solution for $ \mu=0 $ is found by demanding that the term in parenthesis in \eqref{Pxeom} vanishes, which happens for the specific choice $ P=C X^{-1/2} $. The equations of motion vanish identically! Physically one expects higher order corrections to the Lagrangian to become relevant and dictate the dynamics.} $ \ddot\Phi=0 $, i.e. the Spontaneous Symmetry Probing state $ \dot\Phi=\mu $. Remarkably, this solutions exists for any choice of $ P(X) $. Instead, for the driven superfluid, $ \mu\neq 0 $, the equation of motion \eqref{Pxeom} is solved by separation of variables:
\be
\int \rd\dot\Phi \left(  P_{,X}+2XP_{,XX}\right)=- \mu t + C\,.
\ee
For general $ P(X) $, this gives some time dependence $ \Phi(t) $ that is non-linear. For example, the canonical case $  P(X)=X $ gives a quadratic solution $  \ddot \Phi =-\mu $. The low-energy dynamics of perturbations of the driven superfluid is still described by the action \eqref{flatspace}, but interestingly $  d_{n}(t) $ are not constant in time. Rather, they can have a complicated time dependence, which is determined by the equation of state (equivalently by $  P(X)$). Notice that since time translations are non-linearly realized on $  \pi $, this action is not straightforwardly derived using the well-known coset construction for spontaneously broken spacetime symmetries \cite{Ogievetsky}. It would be interesting to find a generalised coset construction that can deal with similar cases. 

Another case captured by our EFT that is not a superfluid is the \textit{imperfect or braided superfluid} discussed in \cite{Pujolas:2011he}. The Lagrangian now contains higher derivative interactions 
\be
\L=P(\partial_{\mu}\Phi\partial^{\mu}\Phi)+G(\partial_{\mu}\Phi\partial^{\mu}\Phi) \Box \Phi\,,
\label{braL}
\ee
for some arbitrary functions $  P $ and $  G $. As in the case of the driven superfluid, the solution of the equation of motion is generically not linear in time $  \Phi\neq t $. As a consequence the effective coefficients $  d_{n} $ and $  \hat d_{n} $ in \eqref{flatspace} do depend on time and must satisfy the constrains \eqref{cond1} and \eqref{cond2}. Finally, notice that when $  \bar\Phi $ is not a linear function, the Goldstone of time translations $  \pi $ does not coincide with the Goldstone boson $  \varphi $ of shift symmetry transformations, as manifest in \eqref{varphi}. We are now ready to tackle gravitational systems.


\section{Shift Symmetry and the EFT of Inflation\label{Sec:SSandEFTofI}}

In this section, we review the EFT of Inflation \cite{Cheung:2007st} and discuss what changes in presence of a shift symmetry. The reader already familiar with the EFT of Inflation may skip directly to Sec.~\ref{Sec:SSandCosmoTime}.


\subsection{Review: the EFT of Inflation\label{Sec:EFTReview}}

In the EFT of Inflation \cite{Creminelli:2006xe,Cheung:2007st}, we assume that the theory consists of the metric $g_{\mu\nu}$ and a single scalar degree of freedom $\Phi$ which takes on a time-dependent profile $\bar{\Phi}(t)$ and sources a Friedmann-Robertson-Walker (FRW) background metric:
\begin{align}
 \rd s^{2}&=-\rd t^{2}+a(t)^{2}\rd \bfx^{2}\ .\label{CosmologicalTimeFRWMetric}
 \end{align} We then study all possible fluctuations about this background, whose forms are constrained by symmetries of the problem, and organize in a derivative expansion.
 
The structure of perturbations is efficiently organized by working in unitary gauge where temporal diffeomorphisms are used to set the perturbation of the scalar field to zero: $\Phi(x)=\bar{\Phi}(t)$ exactly.  In this gauge, the spacelike surfaces of constant $\Phi$ provide a natural ``clock" for the system and the theory is one of the clock and the metric.  In practical terms, this means that the ingredients of the theory are $g_{\mu\nu}$, its associated curvatures and geometric quantities that characterize constant $\Phi$ slices, such as the normal vector $n^{\mu}$, extrinsic curvature $K_{\mu\nu}$, induced metric $\hat{g}_{\mu\nu}=g_{\mu\nu}+n_\mu n_\nu$, intrinsic Ricci curvature $\hat{R}_{\mu\nu}$, etc.  Thus, the Lagrangian takes on the form
 \begin{align}
S&=\int\rd^{4}x\, \sqrt{-g} \, \mathcal{L}\left [g_{\mu\nu},g^{00},R_{\mu\nu\rho\sigma} ,K_{\mu\nu},\nabla_{\mu};t\right ] \ ,\label{GeneralEFTUnitaryGauge}
\end{align}
where any unwritten geometric quantities are redundant with those shown.  After having imposed unitary gauge, the only remaining symmetries of the theory are spatial diffeomorphisms, consistent with the form of \eqref{GeneralEFTUnitaryGauge}.  When expanding the action about a particular FLRW background as in \eqref{CosmologicalTimeFRWMetric}, a similar expression to \eqref{GeneralEFTUnitaryGauge} holds, but with perturbed quantities everywhere.
In discussing the consequences of the shift symmetry in the following we will focus mostly on theories for perturbations in the form
\begin{align}
S &=\int\rd^{4}x \, \sqrt{-g}\, \sum_{n,m=0}^{\infty}\frac{d_{n}{}^{\mu_1\ldots \mu_m}_{\nu_1\ldots\nu_{m}}(t)}{n! \, m!} \left (\delta g^{00}\right )^{n}\delta K^{\nu_{1}}_{\mu _{1}}\ldots \delta K^{\nu_{m}}_{\mu _{m}}\ ,\label{gKAction00}
\end{align}
where $d_{n}{}^{\mu_1\ldots \mu_m}_{\nu_1\ldots\nu_{m}}(t)$ are dimensionful, background time-dependent tensors made of $\delta^\mu_\nu$-functions, which depend on the explicit UV model\footnote{To have an idea of the scaling of the $d_n$'s coefficients, in simple scalar-tensor theories of the type \eqref{braL}, naively assuming one single scale $\Lambda_c$ at play in the Lagrangian, one typically expects $\Lambda_c \sim (d_n)^{1/4}\sim (d_{n}{}^{\mu_1}_{\nu_1})^{1/3}\sim\ldots$ etc. If this were the case, at typical energies of order $H\ll\Lambda_c$ contributions from higher derivative operators are dramatically suppressed by powers of $H/\Lambda_c$. In order to motivate the interest in terms involving not only $(\delta g^{00})^n$ in \eqref{gKAction00}, we point out that there exist well defined theories \cite{Pirtskhalava:2015nla} (see also \cite{Goon:2016ihr} and \cite{Pirtskhalava:2015zwa,Pirtskhalava:2015ebk} for the phenomenological consequences) where e.g. $ (d_{n}{}^{\mu_1}_{\nu_1})^{1/3}\gg  (d_n)^{1/4}$ enhancing the effects of higher derivative operators.} behind the effective description \eqref{gKAction00}. For coefficients of operators linear in $\delta K$, we find it useful to also use the following notation: $d_{n}{}^{\mu}_{\nu}(t)\equiv \hat{d}_{n}(t)\delta^{\mu}_{\nu}$. Generalizations involving, for instance, the Riemann tensor and contractions thereof are straightforward and will not be discussed here.
The effective action contains the following tadpoles\footnote{It is possible to integrate the $\delta K_{\mu}^{\mu}$ tadpole term by parts in order to turn it into a sum of $\sim\left (\delta g^{00}\right )^{n}$ operators \cite{Cheung:2007st}, but we will not perform this manipulation here.}:
\begin{equation}
S_{0}=\int\rd^{4}x \, \sqrt{-g}\, \left [\frac{\Mpl^{2}}{2}R+ d_{0}(t)+d_{1}(t)\delta g^{00}+\hat{d}_{0}(t)\delta K^{\mu}_{\mu}\right ]\label{UniversalEFTAction}\ ,
\end{equation}
where we have defined the metric perturbation relative to the inverse metric $g^{\mu\nu}=\bar{g}^{\mu\nu}+\delta g^{\mu\nu}$. Taking the equations of motion from \eqref{UniversalEFTAction} and demanding that \eqref{CosmologicalTimeFRWMetric} be a solution determines 
\begin{equation}
\Mpl^{2}H^{2}\varepsilon=-d_{1}-\frac{\dot{\hat{d}}_{0}}{2} \ ,\quad 3\Mpl^{2}H^{2}=-d_{0}-2d_{1}+3H\hat{d}_{0} \, .
\label{d0d1Solns0}
\end{equation}
We have labeled our EFT coefficients in such a way as to make the soon-to-be derived recursion relations between different operators particularly compact.  As a result, our notation differs from that of \cite{Cheung:2007st}.  For the reader who is more familiar with the notation of \cite{Cheung:2007st} we summarize in Table \ref{table:notation} the main conversion rules between the two different notations.

\begin{table}[ht]
\caption{Main conversion rules between the different notations for the EFT coefficients.}
\begin{center}
\begin{tabular}{|c|c|c|}
\hline 
Operators in the EFT 											& Coefficients in our notation			& Coefficients in the notation of \cite{Cheung:2007st}				\\
\hline 
$1$ 													& $d_{0}-3H\hat{d}_0-\dot{\hat{d}}_0$ 									& $c-\Lambda$									\\
$\delta g^{00}$ 										& $d_{1}+\frac{1}{2}\dot{\hat{d}}_0$ 									& $-c$											\\
$\left(\delta g^{00}\right){}^{n}\, ,   n\ge 2$ 						& $d_{n}$ 									& $M_{n}^{4}$									\\
$\delta K^{\mu_{1}}_{\nu_{1}}\delta K^{\mu_{2}}_{\nu_{2}}$ 	& $d_{0}{}^{\mu_1\mu_2}_{\nu_1\nu_{2}}$ 	& $-\bar{M}_{2}^{2}\delta^{\mu_{1}}_{\nu_{1}}\delta^{\mu_{2}}_{\nu_{2}}-\bar{M}_{3}^{2}\delta^{\mu_{1}}_{\nu_{2}}\delta^{\mu_{2}}_{\nu_{1}}$	\\
$\delta g^{00}\delta K^{\mu}_{\nu}$ 						& $d_{1}{}^{\mu}_{\nu}$ 							& $-\frac{1}{2}\bar{M}_{1}^{3}\delta^{\mu}_{\nu}$		\\
\hline
\end{tabular}
\end{center}
\label{table:notation}
\end{table}

Differences between models are encoded in the addition of operators which are higher order in $\delta g^{00}$, $\delta K_{\mu\nu}$, $\delta \hat R_{\mu\nu\rho\sigma}$ and the number of derivatives.  For instance, the operators with the fewest derivatives per fluctuation are of the form
\begin{align}
S_{\delta g}=\int\rd^{4}x\sqrt{-g}\, \sum_{n\ge 2}\frac{ d_{n}(t)}{n!}(\delta g^{00})^{n}\ .\label{HigherOrderdeltagAction}
\end{align}
Such operators correspond to the existence of $P(\phi,X)$, $X\equiv -\frac{1}{2}(\nabla\phi)^{2}$, terms in the fundamental Lagrangian, while additions of $\delta K_{\mu\nu}$ would correspond to interactions involving second derivatives $\sim \nabla\nabla\phi$, as has been studied, for example, in Galileon or Horndeski type models \cite{Horndeski:1974wa,Deffayet:2011gz,Langlois:2015cwa,Kobayashi:2010cm,Burrage:2010cu,Gao:2011qe,DeFelice:2013ar}.

As discussed in \cite{Cheung:2007st}, it is convenient  to restore the full diffeomorphism invariance through a St\"uckelberg transformation, acting on $\delta g^{00}$ as 
\be
\delta g^{00} &\rightarrow& - 2 \dot{\pi} - \dot{\pi}^2 + \frac{(\partial_i\pi)^2}{a^2} \, ,\\
\delta {K^i}_{j} &\rightarrow& -\left(\dot H \pi +\frac{1}{2}\ddot{H}\pi^2\right) \delta^i_{j}-(1-\dot{\pi})\partial^i\partial_j\pi \\
&& \quad \quad 	+  \partial^i\dot{\pi}\partial_j\pi + \partial^i\pi\partial_j\dot{\pi} 
	 + H \left(	-  \partial^i\pi\partial_j\pi
	+ \dfrac{1}{2}\delta^i_{j}\partial_k\pi\partial^k\pi\right)
	+ \mathcal{O}(\pi^3)\,. \label{deltaK}
\ee
Under a general diffeomorphism $  x^{\mu}\rightarrow x^{\mu}+\xi^{\mu} $, $  \pi  $ changes as (summarized in Table \ref{tab})
\be
\pi(x)\rightarrow \pi(x+\xi)+\xi^{0}\quad\quad\text{(diffeomorphism)}\,.
\ee
This point of view is especially useful when the mixing between the scalar $\pi$ mode and gravitational degrees of freedom can be neglected.  In this decoupling limit, one finds \cite{Cheung:2007st}:
\begin{align}
S &= \int\rd^{4}x \, \sqrt{-g} \Big[ \frac{\Mpl^{2}}{2}R- \Mpl^{2}\left (3H^{2}(t) +2\dot H(t)\right )+\Mpl^{2}\dot H(t) (\partial\pi)^{2} + \sum_{n=2}^{\infty}\frac{\lambda_n (t)}{n!}\pi^n \label{unitaryg}\\
&\quad+ \sum_{n=2}^{\infty}\frac{d_{n}(t+\pi)}{n!} \left (-2\dot\pi+\partial_{\mu}\pi\partial^{\mu}\pi \right )^{n}+ \sum_{n=1}^{\infty}\frac{\hat d_{n}(t+\pi)}{n!}  \left (-2\dot\pi+\partial_{\mu}\pi\partial^{\mu}\pi \right )^{n} \delta K +\mathcal{O}\left(  \delta K^{2}\right)\Big] \nonumber \,,
\end{align}
where the $\lambda_n (t)$'s are which are determined by the background \cite{Behbahani:2011it}:
\be
\lambda_n (t) =  \Mpl^{2}\left (H\frac{\rd^{n}H}{\rd t^{n}}  - \frac{1}{2}\frac{\rd^{n}\left (H^{2}\right )}{\rd t^{n}} \right )\ .
\ee


\subsection{The EFT of Shift Symmetric Cosmologies\label{Sec:SSandCosmoTime}}

The action \eqref{GeneralEFTUnitaryGauge} descends from a generally covariant action that has simply been evaluated on some background $  \bar \Phi(t) $ in unitary gauge, $\Phi(x)=\bar{\Phi}(t)$.  The symmetries of the gauge fixed action are the subset of symmetries of the original action that also preserve the gauge condition, i.e.$\!$ they are the residual transformations.  For instance, spatial diffeomorphisms are good symmetries of a generic action of the form \eqref{GeneralEFTUnitaryGauge} since they preserve $\Phi(x)=\bar{\Phi}(t)$, whereas temporal diffeomorphisms do not generally preserve the condition.

However, in the limit of an underlying shift symmetry there \textit{does} exist a particular temporal diffeomorphism which preserves the gauge condition, when combined with an internal shift.  Namely, if we perform a diffeomorphism $x^{\mu}\to x^{\mu}+\xi^{\mu}$ where $\xi^{\mu}=c\delta^{\mu}_{0}/\dot{\bar{\Phi}}(t)$ and $c$ an infinitesimal constant, this induces 
\begin{align}
\Phi(x)\to \Phi'(x)=\Phi(t)+c\,,
\end{align}
and the constant $c$ term can be subtracted off by a compensating internal $\Phi(x)\to \Phi(x)-c$ symmetry transformation. Therefore, the consequence of an internal shift symmetry is that the resulting unitary gauge EFT must be invariant under temporal diffeomorphisms of the above form:
\begin{align}
S[g_{\mu\nu},R_{\mu\nu\rho\sigma},K_{\mu\nu},\ldots]&=S[g'_{\mu\nu},R'_{\mu\nu\rho\sigma},K'_{\mu\nu},\ldots]
\end{align}
where $g'_{\mu\nu}=g_{\mu\nu}+\pounds_{\xi}g_{\mu\nu}$ and similar for other quantities. The transformation of $  \pi $ and $  \delta g^{\mu\nu} $ in the full diff-invariant Goldstone action are also easy to find following the St\"uckelberg trick. From the relation $  \Phi(x)=\bar\Phi(t+\pi) $ we deduce the same transformation as in flat space, \eqref{shiftpiflat}, namely $  \Delta \pi=c/\dot{\bar\Phi}(t+\pi) $.


\subsubsection{Clock Time\label{Sec:SSandClockTime}}

The arguments of the previous section can be formulated even more directly by changing the time coordinate.  Rather than working in cosmological time, we can use $\Phi$ directly as the clock.  That is, if we change time coordinates from $t$ to $\phi$ with the relation defined implicitly via\footnote{This, of course, is only possible if we take the standard EFT of Inflation assumption that  $\Phi(t)$ changes monotonically.} $\bar\Phi(t)\equiv \phi$, then the the gauge fixing condition is $\Phi(x^{\mu})=\phi$ where now $x^{\mu}=(\phi,\vec{x})$.  We refer to this choice as ``clock time". The price one pays for using these coordinates is that the background metric takes on the following, somewhat unfamiliar form:
\begin{align}
\rd s^{2}&=-\frac{\rd \phi^{2}}{f(\phi)^{2}}+a(\phi)^{2}\rd \vec{x}^{2}\ ,\label{ClockTimeFRW}
\end{align}
where $f(\phi)=\partial_{t}\bar\Phi(t(\phi))$. In App.~\ref{Appendix:CheckingRInClockTime}, we discuss the form of $\zeta$ in these coordinates both by transforming directly from the comoving gauge expression and by using $\delta N$ arguments.

In clock time, the special temporal diffeomorphism that remains a good symmetry of the gauge fixed action is simply a global time translation: $x^{\mu}\to x^{\mu}+\xi^{\mu}$, $\xi^{\mu}=c\delta^{\mu}_{0}$ with $c$ again a constant.  This produces a simple constant shift of $\Phi$,
\begin{align}
\Phi(x^{\mu})\to \Phi'(x^{\mu})=\Phi(x^{\mu})+c\ ,
\end{align}
which is just the action of the global internal symmetry.  Though these coordinates are unfamiliar, they can be useful for explicit calculations and perhaps provide some conceptual clarity. For instance, in Sec.~\ref{Sec:Sofg00andKmunuRelations} we use clock time to derive EFT recursion relations.

\subsubsection{Goldstone Action in Clock Time\label{subsubsec:ClockTimeGoldstones}}

In this section, we include the form of the Goldstone action calculated in clock time, for completeness. Since by definition $  \partial_{0}\bar \Phi=1 $, the Goldstone of clock-time translations coincides with the Goldstone of shift symmetry transformations and we will hence indicate it by $  \varphi $ to distinguish it from the Goldstone of cosmological-time translations $  \pi $.

For simplicity, we focus only on the lowest derivative interactions \eqref{HigherOrderdeltagAction} such that the Goldstone modes arise from the following terms
\begin{align}
S=\int\rd^{4}x\sqrt{-g}\, \sum_{n\ge 1}\frac{ d_{n}(\phi)}{n!}(\delta g^{00})^{n}\ .\label{DeltaGActionClockTime1}
\end{align}
Following the steps of \cite{Cheung:2007st}, the $\varphi$ field, analogous to $  \pi $ in cosmological time, is introduced by performing a temporal diffeomorphism which in clock time is of the form $\phi\to \phi+\varphi(x)$.  In terms of the $\delta g^{00}$ perturbations, this corresponds to replacing
\begin{align}
\delta g^{00}(x)\to f^{2}(\phi)\left (-2\partial_{\phi}\varphi-(\partial_{\phi}\varphi)^{2}\right )+\frac{(\partial_{i}\varphi)^{2}}{a^{2}(\phi)}+f^{2}(\phi+\varphi)-f^{2}(\phi)\ ,\label{DeltaG00ReplacementClockTime}
\end{align}
where all gravitational perturbations have been neglected, and simultaneously replacing $d_{n}(\phi)\to d_{n}(\phi+\varphi)$.    These steps yield the Lagrangian for scalar fluctuations in clock time:
\begin{align}
S=\int\rd^{4}x\sqrt{-g}\, \sum_{n\ge 1}\frac{ d_{n}(\phi+\varphi)}{n!} \left[  f^{2}(\phi)\left (-2\partial_{\phi}\varphi-(\partial_{\phi}\varphi)^{2}\right )+\frac{(\partial_{i}\varphi)^{2}}{a^{2}(\phi)}+f^{2}(\phi+\varphi)-f^{2}(\phi)\right]^{n}\ ,\label{DeltaGActionClockTime2}
\end{align}
The advantage of clock time is that the shift symmetry for $\varphi$ becomes a simple constant shift (summarized in Table \ref{tab})
\be
\varphi\to \varphi+c\quad\quad\text{(shift symmetry)}\,,
\ee
but the price paid is the somewhat unwieldy form of the action above.  See App.~\ref{App:ClockTimeEFT} for more on the EFT in clock time and App.~\ref{Appendix:CheckingRInClockTime} for an analysis of the comoving curvature perturbation in these coordinates.


\section{Constraints on EFT Parameters\label{Sec:EFTParameterConstraints}}

In this section, we apply the arguments of Sec.~\ref{Sec:SSandEFTofI} to several classes of EFT models.    We start in Sec.~\ref{Sec:Sofg00Relations} by deriving constraints for the simplest set of interactions which have the lowest number of derivatives per fluctuation. Then in Sec.~\ref{Sec:Sofg00andKmunuRelations}, we extend the derivation to theories with higher derivative operators.  Finally, in Sec.~\ref{Sec:SummaryFieldTrans} we collect the transformations of the various perturbations under the actions of interest in a useful reference chart.

\subsection{Relations for $S_{\rm int}[\delta g^{00}]$\label{Sec:Sofg00Relations}}

Consider the following EFT where interactions of the form \eqref{HigherOrderdeltagAction} are added to the universal action \eqref{UniversalEFTAction}  expanded about the cosmological time background \eqref{CosmologicalTimeFRWMetric},
\begin{align}
S&=\int\rd^{4}x\sqrt{-g}\, \left [\frac{\Mpl^{2}}{2}R+\sum_{n=0}^{\infty}\frac{d_{n}(t)}{n!}\left (\delta g^{00}\right )^{n}\right ]\ ,\label{UniversalActionWithdeltag00Terms}
\end{align}
with $d_{0}=\Mpl^2H^2(2\varepsilon-3)$ and $d_{1}=-\Mpl^2H^2\varepsilon$. This theory corresponds to a perfect superfluid, a.k.a.$\!$ a $  P(X) $ theory. It supports many cosmological solutions, such as matter and radiation domination, but it cannot support slow-roll inflation in the shift symmetric limit \cite{Finelli:2017fml}, as we review at the end of this section.

The metric perturbation transforms as $\delta g^{00}\to 	\delta g'^{00}$
 as determined from
\begin{align}
\left (\bar{g}^{\mu\nu}+\delta g^{\mu\nu}\right )&\to \left (\bar{g}^{\mu\nu}+\delta g'^{\mu\nu}\right )=\left (\bar{g}^{\mu\nu}+\delta g^{\mu\nu}\right )+\pounds_{\xi}\left (\bar{g}^{\mu\nu}+\delta g^{\mu\nu}\right )
\end{align}
with $\xi^{\mu}=c\delta^{\mu}_{0}/\dot{\bar{\Phi}}$.  When $\bar{g}^{\mu\nu}$ is the inverse FLRW metric in cosmological time \eqref{CosmologicalTimeFRWMetric}, this corresponds to (summarized in Table \ref{tab})
\begin{align}
\delta g^{00}(t,\bfx)\to\delta g'^{00}(t,\bfx)&=\delta g^{00}(t+\frac{c}{\dot{\bar{\Phi}}},\bfx)-\frac{2c\ddot{\bar{\Phi}}}{\dot{\bar{\Phi}}^{2}}\left (1-\delta g^{00}(t,\bfx)\right )\ ,
\end{align}
understood to hold to $\mathcal{O}(c)$.  Replacing $\delta g^{00}\to \delta g'^{00}$ everywhere in the action and relabeling coordinates, the action changes by
\begin{align}
\Delta S&=\int\rd^{4}x\sqrt{-g}\sum_{n=0}\frac{1}{n!}\left [\frac{2nc\ddot{\bar{\Phi}}}{\dot{\bar{\Phi}}^{2}}d_{n}(t)-\frac{c}{\dot{\bar{\Phi}}}\dot{d}_{n}-\frac{2c\ddot{\bar{\Phi}}}{\dot{\bar{\Phi}}^{2}}d_{n+1}\right ]\left (\delta g^{00}\right )^{n}\label{zerolimit}
\end{align}
and therefore a shift symmetry corresponds to the following relation in these models\footnote{Here we assumed $  \ddot{\bar\Phi}\neq 0 $. To take the $  \ddot{\bar \Phi}\rightarrow 0 $ limit one should go back to the terms in square brackets in \eqref{zerolimit}.}:
\be
\boxed{d_{n+1}=nd_{n}-\frac{\dot{\bar\Phi}}{2\ddot{\bar\Phi}}\dot d_{n}}\,,\label{ShiftSymmetryRelationdeltag00Models}
\ee
for $  n\geq 0 $. So far we have not used the tadpole cancellation conditions \eqref{d0d1Solns0}, a.k.a.$\!$ the background equations of motion.  In Sec.~\ref{Sec:Phenomenology}, this data will be used to solve the relations \eqref{ShiftSymmetryRelationdeltag00Models} recursively;  see e.g.$\!$ \eqref{pxthL}. 

Let us conclude with two final comments. First, from the relations \eqref{ShiftSymmetryRelationdeltag00Models}, it immediately follows that \textit{slow-roll inflation is impossible in $P(X)$ models}.  The speed of sound is determined by $d_{2}$
\begin{align}
\frac{1}{c_{s}^{2}}&=1+\frac{2d_{2}(t)}{\Mpl^{2}H^{2}\varepsilon}\ \label{SpeedOfSoundd2Relation}
\end{align}
and if we use the $n=1$ relation to solve for $\ddot{\bar{\Phi}}$ and substitute the result into the $n=2$ relation, it reduces to
\begin{align}
0&=-\frac{\Mpl^{2}H^{3}}{c_{s}^{2}\dot{\bar{\Phi}}}\varepsilon\left (3+3c_{s}^{2}-2\varepsilon+\eta\right )\ .
\label{beomsst}
\end{align}
Therefore, for stable theories in which $  c_{s}^{2}>0 $, either $\varepsilon=0$, corresponding to the case of ghost inflation \cite{ArkaniHamed:2003uz}, or at least one of $\varepsilon$ and $\eta$ is $\mathcal{O}(1)$, implying that the slow-roll conditions are violated\footnote{Theories of this second type are usually referred to as ``non-attractor'' scenarios, originating with the study of ultra slow-roll inflation \cite{Kinney:2005vj}. See \cite{Bravo:2017gct,Bravo:2017wyw,Cai:2017bxr} for some recent work on such models.}. This result is also derived directly from the $P(X)$ Lagrangian in App.~\ref{App:PofXSection}. It is also worth noticing that, combining the equation $\dot d_0=-2a^{-3}\partial_t(a^3d_1)$ that one derives from the Friedmann equations and with the $n=0$ relation of \eqref{ShiftSymmetryRelationdeltag00Models}, one finds the following differential equation for $\bar \Phi(t)$:
\begin{equation}
\frac{\ddot{\bar{\Phi}}}{\dot{\bar{\Phi}}} = \frac{\partial_t(a^3\dot H)}{a^3\dot H} \, .
\end{equation}
This can be solved exactly:
\begin{equation}
\bar\Phi(t) =\bar \Phi_0 + \bar\Phi_1 \int^t{\rm d}t'  \, a^3(t')\dot H(t') \, ,
\label{PXsolb}
\end{equation}
where $\bar\Phi_0$ and $\bar\Phi_1$ are arbitrary integration constants. We stress that the background metric only determines the solution $\bar{\Phi}(t)$ in the limiting case of $P(X)$ models. We shall see later on that as soon as higher derivative operators are added into the theory this ceases to be true.

Second, the relations \eqref{ShiftSymmetryRelationdeltag00Models} follow from symmetry and are therefore not renormalized to any order in perturbation theory\footnote{In the presence of a Chern-Simon coupling to some gauge sector, there could be non-perturbative corrections that break the shift symmetry and invalidate these relations.}. This is to be contrasted with UV-theories of the form $  P(X)-V(\Phi) $, where the only symmetry breaking source is a potential $V(\Phi)$. Upon inspection, these theories also lead to the relations \eqref{ShiftSymmetryRelationdeltag00Models} for $n\neq 0$  at tree level.  
However, the form of the action $  P(X)-V(\Phi) $ is not protect by any symmetry and loop corrections generate terms $  P(X,\Phi) $ than cannot be written as a separate sum of $  P(X) $ and $  V(\Phi) $. As a consequence, in theories with an action $  P(X)-V(\Phi) $ the relations \eqref{ShiftSymmetryRelationdeltag00Models} are valid (for $  n>0  $) only at tree level and receive corrections at any loop order, which might be small if derivatives of $  V $ are sufficiently suppressed.


\subsection{Relations for $S_{\rm int}[\delta g^{00},\delta K_{\mu\nu}]$\label{Sec:Sofg00andKmunuRelations}}

In this section, we derive the relations between coefficients in the case where the EFT interactions are built not only from $\delta g^{00}$, but also from the higher derivative $\delta K_{\mu\nu}$ operators.  To give a complementary perspective, the relations will be derived using clock time (see Sec.~\ref{Sec:SSandClockTime}) as opposed to cosmological time as in the previous section.

In clock time, the symmetry of the system is a global, constant shift in time: $\phi\to \phi+c$. Let us expand the action in powers of the \textit{full} $g^{00}$ and $K^{\mu}_{\nu}$ (as opposed to perturbations thereof) as
\begin{align}
S_{\rm int}&=\int\rd\phi\rd^{3}\bfx\sqrt{-g}\, \sum_{n,m}\frac{c_{n}{}^{\mu_1\ldots \mu_m}_{\nu_1\ldots\nu_{m}}}{n! \, m!} \left (g^{00}\right )^{n}K^{\nu_{1}}_{\mu _{1}}\ldots K^{\nu_{m}}_{\mu _{m}}\ .\label{gKAction0}
\end{align}
The time translation symmetry now implies that the $c_{n}$ EFT coefficients (suppressing Greek indices) are independent of $\phi$, whereas they would be non-trivial functions of $\phi$ in a generic, non-shift-symmetric theory.

We are usually interested in the action \eqref{gKAction0} expressed as an expansion in the perturbations $\delta g^{00}$ and $\delta K_{\mu\nu}$:
\begin{align}
S_{\rm int}&=\int\rd\phi\rd^{3}\bfx\sqrt{-g}\, \sum_{n,m}\frac{d_{n}{}^{\mu_1\ldots \mu_m}_{\nu_1\ldots\nu_{m}}(\phi)}{n! \, m!} \left (\delta g^{00}\right )^{n}\delta K^{\nu_{1}}_{\mu _{1}}\ldots \delta K^{\nu_{m}}_{\mu _{m}}\ ,\label{gKAction}
\end{align}
where the $d_{n}$'s are now $\phi$-dependent, generically.
Expressions for the $c_{n}$'s in terms of the $d_{n}$'s and background quantities can be straightforwardly derived by making the following replacements in \eqref{gKAction}
\begin{align}
\delta g^{00}&\to g^{00}-\bar{g}^{00}\ , \quad \delta K_{\mu\nu}\to K_{\mu\nu}-\bar{K}_{\mu\nu}\ .
\end{align} and reorganizing the summation.  The consequences of the shift symmetry on the $d_{n}$'s is then simply found by demanding that the $c_{n}$'s are $\phi$ independent:
\begin{align}
0&=\frac{\rd}{\rd \phi}c_{n}{}^{\mu_1\ldots \mu_m}_{\nu_1\ldots\nu_{m}}\left [d_{n},\bar{g}^{00}, \bar{K}_{\mu\nu}\right ]\ .
\end{align}
The result is the following recursive relation:
\begin{equation}
0= \frac{\rd}{\rd\phi}d_{n}{}^{\mu_1\ldots \mu_m}_{\nu_1\ldots\nu_{m}} -
   d_{n+1}{}^{\mu_1\ldots \mu_m}_{\nu_1\ldots\nu_{m}} \frac{\rd \bar g^{00}}{\rd\phi} 
 -  d_{n}{}^{\mu_1\ldots \mu_{m+1}}_{\nu_1\ldots\nu_{m+1}}\frac{\rd}{\rd\phi}  \bar K_{\mu_{m+1}}^{\nu_{m+1}}  \, .
\label{recrelK}
\end{equation}
The analogue relation in the more familiar cosmic time language can be easily derived noticing that, under the time reparametrization $t\rightarrow\phi(t)$, geometric quantities transform in the following way:
\begin{align}
g^{00}&\rightarrow f^{-2}g^{00} \ , \quad N_i= g_{0i}\rightarrow f N_i\ ,\quad
K_{ij}=\frac{1}{2}\sqrt{-g^{00}}\left(\partial_0g_{ij}-D_iN_j-D_jN_i \right)\rightarrow K_{ij}\ ,
 \end{align} 
  where $D_i$ is the covariant derivative written in terms of the induced $3$-metric only. Therefore, since the volume element $\rd^4x\sqrt{-g}$ is invariant, the coefficients of the EFT in the standard cosmic time $t$ are obtained from those in \eqref{gKAction} simply by the rescaling
\begin{equation}
d_{n}{}^{\mu_1\ldots \mu_m}_{\nu_1\ldots\nu_{m}} \rightarrow f^{2n} d_{n}{}^{\mu_1\ldots \mu_m}_{\nu_1\ldots\nu_{m}} \, .
\end{equation}
Plugging this into the recursive relation \eqref{recrelK} and using the definition $f(\phi(t))=1/\dot{\bar \Phi}(t)$, one finds\footnote{Deriving the same relation in cosmological time is straightforward upon the use of the transformation law $\delta K_{\mu}^{\nu}(t,\bfx)\to \delta K'{}_{\mu}^{\nu}(t,\bfx)=\delta K_{\mu}^{\nu}(t,\bfx)+\frac{c}{\dot{\bar{\Phi}}}\dot{\bar{K}}_{\mu}^{\nu}(t)$.}
\begin{equation}
\boxed{  d_{n+1}{}^{\mu_1\ldots \mu_m}_{\nu_1\ldots\nu_{m}}= n d_{n}{}^{\mu_1\ldots \mu_m}_{\nu_1\ldots\nu_{m}}
- \frac{\dot{\bar \Phi}}{2\ddot{\bar \Phi}} \dot{d}_{n}{}^{\mu_1\ldots \mu_m}_{\nu_1\ldots\nu_{m}} 
 + \frac{\dot{\bar \Phi}}{2\ddot{\bar \Phi}} d_{n}{}^{\mu_1\ldots \mu_{m+1}}_{\nu_1\ldots\nu_{m+1}}\dot{\bar K}_{\mu_{m+1}}^{\nu_{m+1}}}  \, .
\label{recrelKt}
\end{equation}
Clearly, in the particular case $d_{n}{}^{\mu_1\ldots \mu_m}_{\nu_1\ldots\nu_{m}}=0$, $\forall m\neq0$, one recovers the result \eqref{ShiftSymmetryRelationdeltag00Models}. As before, these relations are not renormalized to any order in perturbation theory.

In analogy with the procedure in Sec.~\ref{Sec:Sofg00Relations}, the relations \eqref{recrelKt} can be used to simplify the background equations of motion. First, notice that differentiating the second equation in \eqref{d0d1Solns0}, solving for $\dot H$ and plugging it back into the first yields the following background equation
\begin{equation}
\dot d_0-3\dot H\hat{d}_0 = -\frac{2}{a^3}\partial_t(a^3d_1) \, .
\label{eomdKex}
\end{equation}
Requiring shift invariance, namely using the relation with $n=0$ and $m=0$ of \eqref{recrelKt}, the differential equation \eqref{eomdKex} becomes
\begin{equation}
\frac{\ddot{\bar\Phi}}{\dot{\bar{\Phi}}} = \frac{\partial_t(a^3d_1)}{a^3d_{1}} \, ,
\end{equation}
which can be formally solved as
\begin{equation}
\bar\Phi(t) = \bar\Phi_0 + \bar\Phi_1\int^t{\rm d}t'  \, a^3(t')d_1(t') \, ,
\label{soleom}
\end{equation}
for some constant $\bar\Phi_0$ and $\bar\Phi_1$, generalizing Eq. \eqref{PXsolb}. However, as expected, this is not enough to solve unambiguously only in terms of the background metric and derivatives thereof.

\subsection{Summary of Field Transformations\label{Sec:SummaryFieldTrans}}

In Table \ref{tab}, we collect the transformation laws of the various fields of interest under internal shifts and time-diffeomorphisms as a useful reference.  For the $\pi$ and $\varphi$ Goldstone fields, the transformations under actions of an internal shift and a time-diffeomorphisms are independent and therefore listed separately. Shifts act linearly on $  \varphi $ but non-linearly on $  \pi $, while time translations always act non-linearly for non-trivial backgrounds $  \dot{\bar \phi}\neq 0 $. There is always a linear combination of these two symmetries that act linearly on $  \varphi $ (namely $\xi^{0}=-c/\dot{\bar{\Phi}}(t)$), but not on $  \pi $. The only exception happens for $  \dot{\bar \phi} = \text{const} $, in which case $  \pi $ and $ \varphi  $ coincide. At lowest order in derivatives, $  \dot{\bar \phi} = \text{const} $ corresponds to either flat spacetime, $ H=0  $, or the ghost condensate $  P'(X)=0 $ \cite{ArkaniHamed:2003uy}.

In unitary gauge things are slightly different. The transformation law of $\delta g^{00}$ is properly thought of as a combination of the two transformations, and it is therefore listed in the diagonal category in Table \ref{tab}. This fact can again be seen starting from the $\pi$ Goldstone theory by noting that from here we return to unitary gauge by simply setting $\pi\to 0$.  The unitary gauge symmetry transformation can then be seen to arise by starting with the $\pi$ theory and finding the diagonal symmetry under which $\Delta \pi=0$ when evaluated at $\pi=0$, as this is the condition required for preserving unitary gauge.  From Table \ref{tab}, the appropriate action is then simply the shift symmetry plus a diffeomorphism with $\xi^{0}=-c/\dot{\bar{\Phi}}(t)$ and this time-diffeomorphism acts on $\delta g^{00}$ as indicated in the Unitary Gauge column in Table \ref{tab} (up to a sign).

\begin{table}
  \begin{tabular}{r|c | c | c}
   &Unitary Gauge & t-translation Goldstone  & Shift-sym. Goldstone\\ 
   \hline\hline
   Shift Symm.  &  & $ \Delta \delta g^{\mu\nu}=0  $ & $  \Delta \delta g^{\mu\nu}=0 $ \\ 
   &  & $\Delta \pi=\frac{c}{\dot{\bar\Phi}(t+\pi)}  $ & $ \Delta \varphi=c $ \\ \hline
   Diagonal   &$ \Delta \delta g^{00}=\frac{c}{\dot{\bar{\Phi}}} \dot{\delta g^{00}}-\frac{2c\ddot{\bar{\Phi}}}{\dot{\bar{\Phi}}^{2}}\left (1-\delta g^{00}\right )  $ & & \\ \hline
   t-diff. & &  $ \Delta \delta g^{\mu\nu}=\pounds_{\xi}g^{\mu\nu} $ & $  \Delta  \delta g^{\mu\nu}=\pounds_{\xi}g^{\mu\nu} $ \\
   & &  $  \Delta \pi=\xi^{0}\dot\pi+\xi^{0}  $ & $   \Delta \varphi=\xi^{0}\dot\varphi+\xi^{0}\dot{\bar{\Phi}}(t) $ \\ \hline
  \end{tabular}
\caption{This table summarizes the transformations of the time-translation Goldstone $  \pi $, the shift symmetry Goldstone $  \varphi $ and the metric under time-diffeomorphisms and the shift symmetry.  Cosmological time is used in all of the above cases.\label{tab}}
\end{table}

\section{Shift Symmetric Adiabatic Mode\label{Sec:AdiabaticModes}}

In this section, we discuss the adiabatic mode for the comoving curvature perturbation $\zeta$ which descends from the shift symmetry.  The naive transformation for $\zeta$ is a symmetry for the scalar action when written in terms of $\zeta$, the lapse and the shift.  However, it fails to be a symmetry once these constraint fields are integrated out.  The essential reason for this discrepancy is that the shifted profiles do not preserve the solution to the constraint equations, which is what is used to remove the lapse and shift from the action.  By following Weinberg's adiabatic mode argument \cite{Weinberg:2003sw}, we derive the corrected $\zeta$-shift, which is a true symmetry of the $\zeta$-action. This adiabatic mode leads to Ward identities amongst cosmological correlators, as we discussed in \citep{Finelli:2017fml}.

 We now discuss the details of the adiabatic mode, expanding upon the results presented in \cite{Finelli:2017fml,Pajer:2017hmb}.  We will restrict our discussion to the case of EFTs of the form considered in Sec.~\ref{Sec:Sofg00Relations}:
\begin{align}
S&=\int\rd^{4}x\sqrt{-g}\, \left [\frac{\Mpl^{2}}{2}R+\sum_{n=0}^{\infty}\frac{d_{n}(t)}{n!}\left (\delta g^{00}\right )^{n}\right ]\ .\label{UniversalActionWithdeltag00Terms2}
\end{align} This is done for concreteness and simplicity; generalization to other cases is straightforward.

Consider the action for the comoving curvature perturbation $\zeta$ introduced as a metric perturbation in the following form
\begin{align}
\rd s^{2}&=-\left (1+\delta N\right )^{2}\rd t^{2}+a(t)^{2}e^{2\zeta}\delta_{ij}\left (\rd x^{i}+\partial^{i}\psi\rd t\right )\left (\rd x^{j}+\partial^{j}\psi\rd t\right )\ ,\label{ComovingMetricPerturbations}
\end{align}
where we are working in comoving gauge and have only included scalar perturbations.  Above, and in what follows, spatial indices will be raised and lowered with the flat $\delta_{ij}$ metric.  Plugging the decomposition \eqref{ComovingMetricPerturbations} into \eqref{UniversalActionWithdeltag00Terms2}, the quadratic terms are found to be:
\begin{align}
S^{(2)}&=\int\rd t\rd^{3}\tilde{k}\, \Mpl^{2}a(t)^{3}\Bigg[-3\dot{\zeta}_{\bfk}\dot{\zeta}_{-\bfk}-k^{2}\left (\psi_{\bfk}\dot{\zeta}_{-\bfk}+\psi_{-\bfk}\dot{\zeta}_{\bfk}\right )+3H\left (\delta N_{\bfk}\dot{\zeta}_{-\bfk}\delta N_{-\bfk}\dot{\zeta}_{\bfk}\right )\nn
&\quad+k^{2}H\left (\delta N_{\bfk}\delta N_{-\bfk}+\delta N_{\bfk}\psi_{-\bfk}+\delta N_{-\bfk}\psi_{\bfk}\right )	+k^{2}\left (\zeta_{\bfk}\zeta_{-\bfk}+\delta N_{\bfk}\zeta_{-\bfk}+\delta N_{-\bfk}\zeta_{\bfk}\right )/a^{2}\nn
&\quad-H^{2}\left (1-\varepsilon\right )\delta N_{\bfk}\delta N_{-\bfk}-\frac{H^{2}\varepsilon(c_{s}^{2}-1)}{c_{s}^{2}}\delta N_{\bfk}\delta N_{-\bfk}\Bigg]\ ,\label{ZetaActionBeforeIntegratingOutConstraints}
\end{align}
where we have symmetrized over momentum labelings and replaced $d_{2}$ by its expression in terms of the speed of sound \eqref{SpeedOfSoundd2Relation}.

From the arguments of Sec.~\ref{Sec:EFTParameterConstraints}, the underlying shift symmetry turns into the symmetry transformation $g_{\mu\nu}\to g_{\mu\nu}+\pounds_{\xi}g_{\mu\nu}$ with $\xi^{\mu}=c\delta^{\mu}_{0}/\dot{\bar{\Phi}}$.  In terms of the perturbations in \eqref{ComovingMetricPerturbations}, this corresponds to:
\begin{align}
\zeta_{\bfk}&\to \zeta_{\bfk}+c\frac{H}{\dot{\bar{\Phi}}}\tilde{\delta}^{3}(\bfk)\ , \quad
\delta N_{\bfk}\to \delta N_{\bfk}-c\frac{\ddot{\bar{\Phi}}}{\dot{\bar{\Phi}}^{2}}\tilde{\delta}^{3}(\bfk)\ , \quad \psi_{\bfk}\to \psi_{\bfk}\label{MetricPerturbationSymmetriesNaive}\ .
\end{align}
If we substitute \eqref{MetricPerturbationSymmetriesNaive} into \eqref{ZetaActionBeforeIntegratingOutConstraints}, the change to the action is
\begin{align}
\Delta S^{(2)}&=c\int\rd t\, \Mpl^{2}a(t)^{3}\left[-\frac{2H^{2}\varepsilon\left (3Hc_{s}^{2}\dot{\bar{\Phi}}+\ddot{\bar{\Phi}}\right )}{c_{s}^{2}\dot{\bar{\Phi}}^{2}}\delta N_{\bfk}+\frac{6H^{2}\varepsilon\left [\left (-3+2\varepsilon-\eta\right )H\dot{\bar{\Phi}}+\ddot{\bar{\Phi}}\right ]}{\dot{\bar{\Phi}}^{2}}\zeta_{\bfk}\right]\ ,
\end{align}
after integrating by parts.  In a generic theory, the above is non-zero, but in an shift symmetric theory the recursion relations \eqref{ShiftSymmetryRelationdeltag00Models} ensure that both coefficients vanish.

However, if we integrate out the $\psi$ and $\delta N$ constraint fields to build the action for $\zeta$ alone, the naive transformation \eqref{MetricPerturbationSymmetriesNaive} is \textit{not} a symmetry of the resulting action.  The constraint equations which arise as the $\delta N$ and $\psi$ EOM are
\begin{align}
0&=k^{2}\left (\zeta_{\bfk}/a^{2}+H\psi_{\bfk}\right )+\frac{\varepsilon H\dot{ \zeta}_{\bfk}}{c_{s}^{2}}\ , \quad 0=k^{2}\left (H\delta N_{\bfk}-\dot{\zeta}_{\bfk}\right )\ , \label{ConstraintEquationsFromLapseShift}
\end{align}
which has the following solutions at non-trivial $\bfk$:
\begin{align}
\delta N_{\bfk}&=\frac{\dot{\zeta}_{\bfk}}{H}\ , \quad \psi_{\bfk}=-\frac{\zeta_{\bfk}}{a^{2}H}-\frac{\varepsilon \dot{\zeta}_{\bfk}}{c_{s}^{2}k^{2}}\ .\label{ConstraintSolutions}
\end{align}
Substituting \eqref{ConstraintSolutions} into \eqref{ZetaActionBeforeIntegratingOutConstraints}, one is left with the standard quadratic $c_{s}$ $\zeta$-action:
\begin{align}
S^{(2)}_{\zeta}&=\int\rd t\rd^{3}\tilde{k} \, \Mpl^{2}a(t)^{3}\varepsilon\left [c_{s}^{-2}\dot{\zeta}_{\bfk}\dot{\zeta}_{-\bfk}-\frac{k^{2}}{a^{2}}\zeta_{\bfk}\zeta_{-\bfk}\right ]\ .	\label{ZetaActioncs}
\end{align}
Now, if the naive $\zeta_{\bfk}\to \zeta_{\bfk}+c\frac{H}{\dot{\bar{\Phi}}}\tilde{\delta}^{3}(\bfk)$ transformation is applied to \eqref{ZetaActioncs} , the variation of the action, $\Delta S^{(2)}_{\zeta}$, is a complicated sum of terms which does not vanish upon use of the recursion relations \eqref{ShiftSymmetryRelationdeltag00Models}.  Equivalently, it can be checked that the field profile corresponding to the naive shift, $\zeta_{\bfk}=\frac{H}{\dot{\bar{\Phi}}}\tilde{\delta}^{3}(\bfk)$, does not satisfy the $\zeta$ equations of motion.

The failure of the $\zeta$-transformation to be a symmetry of \eqref{ZetaActioncs} arises from the fact that the shifts \eqref{MetricPerturbationSymmetriesNaive} do not preserve the form of the solutions \eqref{ConstraintSolutions} that were used to integrate out the constraints.  In practical terms, this means that  in \eqref{ZetaActioncs} we have replaced, for instance, $\delta N_{\bfk}\to \dot{\zeta}/H$ everywhere and whereas the factors of $\delta N_{\bfk}$ in the initial action \eqref{ZetaActioncs} shifted as $\delta N_{\bfk}\to \delta N_{\bfk}-c\frac{\ddot{\bar{\Phi}}}{\dot{\bar{\Phi}}^{2}}\tilde{\delta}^{3}(\bfk)$, the replacement factors of $\dot{\zeta}/H$ in \eqref{ZetaActioncs} now shift instead as\footnote{The combination $H\delta N_{\bfk}-\dot{\zeta}_{\bfk}$ shifts as $H\delta N_{\bfk}-\dot{\zeta}_{\bfk}\to H\delta N_{\bfk}-\dot{\zeta}_{\bfk}+c\frac{\varepsilon H^{2}}{\dot{\bar{\Phi}}}\tilde{\delta}^{3}(\bfk)$ which still trivially satisfies \eqref{ConstraintEquationsFromLapseShift} as the shift only has support at $\bfk=0$.}
\begin{align}
 \dot{\zeta}_{\bfk}/H\to \dot{\zeta}_{\bfk}/H+\frac{c}{\dot{\bar{\Phi}}^{2}}\left (\ddot{\bar{\Phi}}-\varepsilon H\dot{\bar{\Phi}}\right )\tilde{\delta}^{3}(\bfk)\ .
 \end{align}
 This is the reason why the $\zeta$ action \eqref{ZetaActioncs} is not invariant under the naive transformation in \eqref{MetricPerturbationSymmetriesNaive}.

 Weinberg's adiabatic mode construction \cite{Weinberg:2003sw} instructs us how to cure this issue.  The procedure is to improve the diffeomorphism that generated \eqref{MetricPerturbationSymmetriesNaive} by adding to it an additional residual diffeomorphism that ensures that the relations \eqref{ConstraintEquationsFromLapseShift} are preserved.  The full ``adiabatic" diffeomorphism that preserves the constraints turns out to be a combination of the naive temporal shift plus a time-dependent dilation \cite{Finelli:2017fml}:
 \begin{align}
  \xi^{\mu}_{\rm adiabatic}&=c\left (\frac{1}{\dot{\bar{\Phi}}},x^{i}\int^{t}\rd t'\, \frac{\varepsilon H^{2}}{\dot{\bar{\Phi}}}\right )\label{AdiabaticDiffeomorphism}\ .
  \end{align} 
  The total, improved transformation is then 
  \begin{align}
  \zeta_{\bfk}\to\zeta_{\bfk}+c\left (\frac{H}{\dot{\bar{\Phi}}}+\int^{t}\rd t'\, \frac{\varepsilon H^{2}}{\dot{\bar{\Phi}}}\right )\tilde{\delta}^{3}(\bfk)\ ,\label{ImprovedZetaTransformation}
  \end{align} 
  to lowest order in $\zeta$,  and it is straightforward to check that \eqref{ImprovedZetaTransformation} is a symmetry of \eqref{ZetaActioncs} and, correspondingly, that the associated $\zeta_{\bfk}$ profile solves the equations of motion stemming from \eqref{ZetaActioncs}.
  
  Finally, we note that the situation here is somewhat different than the more familiar cases of the dilation and special conformal residual symmetries of cosmological spacetimes, as the adiabatic correction term in \eqref{AdiabaticDiffeomorphism} already affects the  field-independent part of $\zeta$'s transformation law, i.e. the ``non-linear" term.  For instance, a similar adiabatic analysis is carried out in \cite{Hinterbichler:2013dpa} for the residual special conformal diffeomorphisms where it was found that the adiabatic correction term only affects $\zeta$'s transformation at linear order in $\zeta$.  Specifically, the naive special conformal diffeomorphism is purely spatial,
  \begin{align} 	
  \xi^{i}_{\rm naive\ SCT}=2({\bf b}\cdot \bfx)x^{i}	-x^{2}b^{i}\ ,
  \end{align}
  and the compensating piece which is required to create a constraint-solution-preserving adiabatic mode is a time-dependent translation:
  \begin{align}
\xi^{i}_{\rm adiabatic \ SCT}&=2({\bf b}\cdot \bfx)x^{i}	-x^{2}b^{i}-2b^{i}\int^{t}\frac{\rd t'}{H}\ .\label{SCTAdiabaticDiff}
  \end{align}
  At lowest order, the corresponding symmetry acts on $\zeta$ in position space as
  \begin{align}
  \zeta\to \zeta+H\xi^{0}+\frac{1}{3}\partial_{i}\xi^{i}\ ,
  \end{align}
and hence the additional term in \eqref{SCTAdiabaticDiff} does not affect the non-linear term of the transformation law.  There is trivially no adiabatic correction to $\zeta$'s symmetry in the dilation case, as the naive dilation already corresponds to a proper adiabatic mode.

\section{Interactions and Non-Gaussianity\label{Sec:Phenomenology}}

In this section, we discuss the strength of interactions in the shift symmetry effective theory \eqref{UniversalActionWithdeltag00Terms} and estimate the size of the associated non-Gaussianity. For simplicity, we consider only interactions at leading order in derivatives, i.e. those of the form $\sim	\left (\delta g^{00}\right )^{n}$.

\subsection{Goldstone Interactions}

We start by deriving the form of the interactions in terms of $\pi$, the Goldstone of time-translation breaking.  The action is then used to estimate the size of non-Gaussianity produced by the $\sim \pi\left (\partial_{i}\pi\right )^{2}$ operator which is typically ignored in EFT of inflation analyses, but can be non-negligible in shift-symmetric theories.

At leading order in derivatives, one can express all effective coefficients in terms of the Hubble parameter and its derivatives (a.k.a.$\!$ the slow roll parameters) using \eqref{ShiftSymmetryRelationdeltag00Models}. Intuitively, this is to be expected given our discussion in Sec.~\ref{Sec:Introduction}: in a small enough region, perturbations can be thought of as shifts back and forth along the background solution and so their interactions are fixed by the Taylor expansion of background quantities such as the Hubble parameter. Indeed, up to cubic order in perturbations, the effective Lagrangian \eqref{UniversalActionWithdeltag00Terms} is
\be
\mathcal{L} &=& \frac{\Mpl^2}{2}R - \Mpl^2 \left(3H^2 +2 \dot{H} \right) 
+ \Mpl^2\dot{H}\delta g^{00} + \frac{\Mpl^2\dot{H}}{4}\frac{c_s^2-1}{c_s^2}(\delta g^{00})^2
\label{pxthL}\\
&&\hspace{2cm}+ \frac{\Mpl^2\dot{H}}{6c_s^2} \left[ c_s^2-1
-\frac{H\dot{\bar{\Phi}}}{\ddot{\bar{\Phi}}}\frac{ 2s + (1-c_s^2)(2\varepsilon-\eta) }{4}
\right](\delta g^{00})^3 
+ \ldots\,, \nonumber
\ee
where $\ddot{\bar{\Phi}}/(H\dot{\bar{\Phi}})=-3c_s^2$ if the shift symmetry is exact\footnote{Otherwise, for $  P(X,\Phi) $ theories, the background equations of motion imply
\be
\frac{\ddot{\bar{\Phi}}}{H\dot{\bar{\Phi}}}=-c_s^2 \left( 3-\frac{\dot{\bar \Phi}(P_{,\Phi}-2\dot{\bar \Phi}^2P_{,X\Phi})}{2\Mpl^2H^3\varepsilon} \right)\,.\label{nuova}
\ee
We stress that for theories of the type $\mathcal{L}=P(X)-V(\Phi)$ where the potential is the only symmetry breaking source Eq. \eqref{pxthL} is formally identical, but with $\ddot{\bar{\Phi}}/(H\dot{\bar{\Phi}})$ given by \eqref{nuova} and with $P_{,\Phi}\equiv-V_{,\Phi}$. Indeed, as already mentioned, at tree level the recursive relations \eqref{ShiftSymmetryRelationdeltag00Models} are unaffected except for the one with $n=0$.}.
The coefficients of higher order $(\delta g^{00})^n$, $n>3$ operators are also determined by the recursive relations \eqref{ShiftSymmetryRelationdeltag00Models}.  For the reader familiar with the notation of \cite{Senatore:2009gt,Ade:2015ava}, our $d_{3}$ coefficient can be related to their $\tilde{c}_{3}$ via
\begin{equation}
\tilde{c}_3 (c_s^{-2}-1) = \frac{2c_s^2d_3}{\Mpl^2\dot{H}} 
=  2(c_s^2-1) \left[ 1
+\frac{H\dot{\bar{\Phi}}}{4\ddot{\bar{\Phi}}} \left(  \frac{2s}{1-c_s^2} + 2\varepsilon-\eta \right)
\right] \, ,
\end{equation}
and when the shift symmetry is exact, this can be re-expressed as
\begin{equation}
\tilde{c}_3\left(c_s^{-2}-1 \right) 
= \frac{1}{2}\left(3c_s^2 - 4 + c_s^{-2} \right) + \frac{s}{3c_s^2} \, ,
\end{equation}
where $s\equiv \dot{c}_{s}/(c_{s}H)$.

For phenomenology it is often convenient to make the scalar degree of freedom explicit using the St\"uckelberg trick.  The quadratic action reads 
\begin{align}
  S^{(2)}_{\pi}&=\int\rd^{4}x\, \sqrt{-g} \frac{\varepsilon\Mpl^{2}H^{2}}{c_{s}^{2}}\left [\dot{\pi}^{2}-c_{s}^{2}\frac{(\partial_{i}\pi)^{2}}{a^{2}}-3c_s^2H^2\varepsilon\pi^2\right ] \, .
  \end{align}
 It proves useful to restore a sort of fake Lorentz invariance by rescaling the spatial coordinates as $x^i\rightarrow \tilde x^i \equiv  x^i/c_s$ \cite{Baumann:2011su}. Performing this relabeling and additionally canonically normalizing the scalar field as
\begin{equation}
\pi_c \equiv \pi \sqrt{2c_sM_\text{Pl}^2H^2\varepsilon} \equiv \pi f_\pi^2 \, ,
\end{equation}
the ultimate quadratic action is
\begin{equation}
S^{(2)}_{\pi_c} = \frac{1}{2}\int \D t \, \D^3\tilde{x}\left[ (\tilde{\partial}_\mu\pi_c)^2 - m^2\pi_c^2 \right] \, ,
\label{St2accan}
\end{equation}
where
\begin{equation}
m^2 = \frac{H^2}{4}\left[-5s^2+8\varepsilon^2+s(6+6\varepsilon-4\eta) + \eta(6+\eta) -2\varepsilon(6+6c_s^2+5\eta) +2 \frac{\dot{s}+\dot{\eta}}{H} \right] \ .
\label{mcanS}
\end{equation}
Carrying out the same steps for the cubic action results in the following:
\begin{multline}
S^{(3)}_{\pi_c} = \int \D t \, \D^3\tilde{x} \bigg[
\frac{c_s^2-1}{2 \Lambda_\star^2}\dot{\pi}_c\frac{(\tilde \partial_i\pi_c)^2}{a^2}
+ \frac{(1-c_s^2)(2\tilde c_3 + 3c_s^2)}{6\Lambda_\star^2} \dot{\pi}^3_c 
\\
+ \frac{(1-c_s^2)s-(1-3c_s^2)(2\varepsilon-\eta)}{4 \Lambda_\star^2}
H \pi_c\frac{(\tilde \partial_i\pi_c)^2}{a^2}
\\
+ \frac{c_s^2\left[(3c_s^2-7)s+(1-3c_s^2)(2\varepsilon-\eta)\right]-2\tilde c_3(1-c_s^2)(s-2\varepsilon+\eta)}{4 \Lambda_\star^2}
H \pi_c\dot \pi_c^2
+ \frac{H^3}{\Lambda_\star^2}C_{\pi_c^3} \pi_c^3 \, ,
\label{St3accan}
\end{multline}
where we defined the scale
\begin{equation}
\Lambda_\star \equiv c_s f_\pi \gg H\,, \label{LambdaStar}
\end{equation}
which, roughly, controls the strength of the derivative operators. In \eqref{St3accan} the dimensionless coefficient $C_{\pi_c^3}$ is a lengthy combination of slow-roll parameters that we omit for simplicity.

Finally, we briefly estimate the relative size of the non-Gaussianity induced by the $\pi_c(\tilde{\partial}_i\pi_c)^2$ operator.  Because an approximate constant shift symmetry is typically assumed for $\pi_{c}$, this operator is not usually considered as its coefficient is taken to be slow-roll suppressed.  However, this need not be the case in general. Comparing its contribution to $f_{\text{NL}}$ against that of $\dot \pi_c(\tilde{\partial}_i\pi_c)^2$, for example, yields
\begin{equation}
\frac{f_{\text{NL}}^{\dot \pi_c(\tilde{\partial}_i\pi_c)^2}}{f_{\text{NL}}^{ \pi_c(\tilde{\partial}_i\pi_c)^2}}
\sim \frac{1-c_s^2}{(1-c_s^2)s-(1-3c_s^2)(2\varepsilon-\eta)} \, ,
\label{compfNL}
\end{equation}
which in principle can be $\sim\mathcal{O}(1)$ if there is for instance some strong time dependence in the parameters such that the denominator turns out not to be parametrically slow roll suppressed.

\subsection{Strong Coupling}

In this section we estimate the strong coupling scale of the shift-symmetric theory.  We find that imposing a shift symmetry does not change the usual estimate for this scale.

It is often assumed that the action for $\pi$ has an approximate constant shift symmetry, $\pi_{c}\to \pi_{c}+c$. As a result, operators with undifferentiated factors of $\pi$ are typically ignored, as their coefficients are taken to be slow-roll suppressed.  For a generic shift-symmetric theory, this is not generally true. Indeed, in addition to the standard operators $\dot{\pi}_c^3$ and $\dot{\pi}_c(\tilde{\partial}_i\pi_c)^2$ discussed in \cite{Cheung:2007st,Senatore:2009gt}, in \eqref{St3accan} there is also a $\pi_c(\tilde{\partial}_i\pi_c)^2$ term whose coefficient is not generally small, for instance. This is to be expected because the shift symmetry in cosmic time, $ \Delta \pi=c/\dot{\bar\Phi}(t+\pi)   $, does not in general correspond to a simple constant shift. When such lower derivative operators also exist in the effective theory, it's possible that they  can play a non-trivial role in determining the theory's strong coupling scale.  We repeat the standard strong coupling scale estimate in the presence of this additional operators now.

It follows from dimensional analysis that a generic operator at order $\mathcal{O}(\pi_c^n)$ is of the form
\begin{equation}
S^{(n)}_{\pi_c} \supset \int \D t \, \D^3\tilde{x} \, C_{n,m,k}(\varepsilon, \eta,  c_s, s, \ldots) \frac{\pi_c^{m}\dot{\pi}_c^{n-m-2k}(\tilde{\partial}_i\pi_c)^{2k}}{\Lambda_\star^{2n-4}H^{-m}} \, ,
\end{equation}
where the dimensionless coefficients $C_{n,m,k}(\varepsilon, \eta,  c_s, s, \ldots)$ are generically functions of the indicated arguments.  In the standard EFT of Inflation, where $\pi_{c}$ is always differentiated, these coefficients are taken to be negligible if $m\neq 0$ and the strong coupling scale is simply $\Lambda_{\star}$. In our scenario, where the coefficients can be non-negligible when $m\neq 0$, there is a second scale $H$ and the energy associated to the above operator is:
\begin{equation}
E_{n,m,k} \sim \Lambda_\star \,  C_{n,m,k}^{-1} \left(\frac{\Lambda_\star}{H} \right)^{\frac{m}{2n-m-4}} \, .
\end{equation}
Clearly, $E_{n,m,k}$ increases with $m$, unless $2n-m-4<0$.  However, the only operator that satisfies this condition is a relevant cubic $\propto \pi_{c}^{3}$ operator, corresponding to $n=m=3$, $k=0$, which does not play a role in the following discussion. Therefore, if $C_{n,m,k}(\varepsilon, \eta,  c_s, s, \ldots)$ are all of the same order of magnitude, then the strong coupling scale is generically fixed by the standard derivative operators with $m=0$, i.e.~the scale is still $\Lambda_{\star}$. 

For some specific cubic operators, we can be more precise about the form of the $C_{n,m,k}$'s, which allow us to discuss the theory's cutoff in sharper terms.  In \cite{Senatore:2009gt} the perturbative unitarity cutoff induced by $\dot \pi_c(\tilde{\partial}_i\pi_c)^2$ and $\dot \pi_c^3$ were found to scale as
\begin{equation}
\Lambda_{\dot \pi_c(\tilde{\partial}_i\pi_c)^2} \sim \frac{\Lambda_\star}{(1-c_s^2)^{1/2}}
\, , \qquad
\Lambda_{\dot \pi_c^3} \sim \frac{\Lambda_\star}{(1-c_s^2)^{1/2}(c_s^2+3\tilde{c}_3/3)^{1/2}} \, ,
\label{scanold}
\end{equation}
where we are dropping $4\pi$-factors for simplicity.
An analogous computation can be done also for the other operators in the cubic action \eqref{St3accan}. For instance,
\begin{equation}
\Lambda_{\pi_c(\tilde{\partial}_i\pi_c)^2} \sim \frac{\Lambda_\star}{(1-c_s^2)s-(1-3c_s^2)(2\varepsilon-\eta)} \frac{\Lambda_\star}{H} \, .
\label{scanw}
\end{equation}
Notice that in the case of standard slow roll evolution, $\varepsilon, \eta, s\ll1$, the scale \eqref{scanw} is expected to be parametrically higher than \eqref{scanold}, which therefore can be used as an estimation of the unitarity cutoff in the effective theory. By contrast, in the presence of some ``strong'' time dependence, e.g. $\eta, s\lesssim\mathcal{O}(1)$, as required for instance to solve the background equations of motion \eqref{beomsst} in the shift symmetric theory \eqref{St3accan}, the pre-factor in \eqref{scanw} does not provide necessarily a huge enhancement, even if, on the other hand, the ratio $\Lambda_\star/H\gtrsim 1$ still tends to push the strong coupling associated with $\pi_c(\tilde{\partial}_i\pi_c)^2$ above $\Lambda_\star$. Nevertheless, it seems that for a speed of sound close enough to $1$, i.e. $(1-c_s^2)^{1/2}\lesssim H/\Lambda_\star$, the scale \eqref{scanw} could become in principle smaller than \eqref{scanold} found in \cite{Cheung:2007st,Senatore:2009gt}, determining the energy at which unitarity breaks down. However, in order to avoid fine tuning, operators with less derivatives in \eqref{St3accan} can not generically become strongly coupled at a scale that is lower than the one associated with $\dot \pi_c(\tilde{\partial}_i\pi_c)^2$ and $\dot \pi_c^3$. Indeed, requiring for instance that quantum corrections to the couplings of the derivative operators $\dot \pi_c(\tilde{\partial}_i\pi_c)^2$ and $\dot \pi_c^3$, due to loop diagrams involving e.g. the operators $\pi_c(\tilde{\partial}_i\pi_c)^2$ and $\pi_c\dot \pi_c^2$, are at most of the same order of their tree level values, one finds the condition $(1-c_s^2)^{1/2}\gtrsim H/\Lambda_\star$.
As a result, one can still rely on \eqref{scanold} as an estimation of the energy scale at which additional UV degrees of freedom are expected to come into play.



\section{Discussion\label{Sec:Discussion}}

In this paper we have synthesized two concepts central to the study of fundamental cosmology: shift symmetries of scalar fields, often invoked to explain the naturalness of putative inflationary potentials, and the EFT of Inflation \cite{Cheung:2007st}, which has become the standard framework for studying cosmological perturbations in a united way. 
Primarily, we have derived the constraints on EFT coefficients that are imposed by an exact, fundamental shift symmetry.  These take the form of an infinite tower of recursion relations that govern the interactions of the resulting Goldstone theory, whose symmetry properties we have also elucidated.

There are various avenues along which to extend the present work.  Two interesting lines of research are as follows:
\begin{itemize}
\item One motivation for this paper was to understand how internal symmetries and breaking patterns encode themselves into the EFT of Inflation.  The present case was but the simplest example and one may consider other breaking patterns involving different additional (internal, gauge, spacetime) symmetries, which, in the spirit of \cite{Nicolis:2015sra}, could be also thought of as particular realizations of different forms of ``matter'' (see also \cite{Bartolo:2015qvr,Lin:2015cqa}). As studied in Sec.~\ref{Sec:AdiabaticModes} and \cite{Finelli:2017fml}, adiabatic modes are associated to the internal symmetries and it would be interesting to work out their forms in these alternatives scenarios.

\item Exact global symmetries are an idealization and are argued to not exist in consistent gravitational theories \cite{ArkaniHamed:2006dz,Kallosh:1995hi,Banks:2010zn}.  Indeed, in the context of inflation, we do not expect a perfect, global shift symmetry to be realistic, both for the aforementioned theoretical considerations and due to the practical issue that our recursive relations are in tension with the existence of slow-roll backgrounds, at least in the simplest shift-symmetric models.  It would therefore be interesting to understand how soft symmetry breaking terms affect the conclusions of this work.  Such a study would again be profitably carried out in EFT language, in which many classes of softly-broken models could simultaneously be analyzed.  Analogous to the psuedo-Goldstone boson analysis of QCD in which the squared pion masses are linked to the symmetry-breaking quark mass terms, we expect that this line of work would give an interpretation of cosmological observables in terms of the scale characterizing the breaking of the shift symmetry\footnote{This scale is conceptually distinct from the scale associated to the breaking of \textit{time}-translations, as studied in, e.g., \cite{Baumann:2011ws}.}. Finally, it would be particularly interesting to use this framework to determine whether there is some minimal scale at which the shift-symmetry must be broken in order to be consistent with a healthy inflationary background. These questions are left for future work \cite{InPreparation}.

\end{itemize} 

\vspace{1cm}

\textbf{Acknowledgements:}  It is our pleasure to thank Paolo Creminelli, Sadra Jazayeri and Federico Piazza for comments. B.F.,
G.G. and E.P. are supported by the Delta-ITP consortium, a program of the Netherlands organization for scientific
research (NWO) that is funded by the Dutch Ministry of Education, Culture and Science (OCW). L.S. is  supported by the Netherlands organization for scientific research (NWO).


\appendix


\section{Noether Current}

Because the diffeomorphism generated by $\xi^\mu = \delta^\mu_0/\dot{\bar\Phi}$ is a symmetry of the action, the Lagrangian must transform as a scalar, $\Delta \L = \xi^\mu \nabla_\mu \L$. Together with the transformation of the metric determinant, the action changes by a boundary term:

\be
\Delta S = \int \rd^4x \sqrt{-g}\,\nabla_\mu (\xi^\mu \L),
\ee
so that the structure current for the associated conserved current is $k^\mu = \xi^\mu \L$. Meanwhile, we can write the variation of the action in terms of the variations of $g^{00}$ and $K_{\mu\nu}$. The transformation of $g^{00}$ is known. As for $K_{\mu\nu}$, it is not a covariant object in the unitary gauge we are using, so it will not, under a general diffeomorphism, transform as a tensor--this is why such transformations are not symmetries to begin with. However, the specific diffeomorphism $\xi^\mu$ we are considering is a symmetry, and thus the extrinsic curvature must transform covariantly under it if the overall Lagrangian is to be a scalar. Thus the action varies as:

\be
\Delta S = \int \rd^4 x\sqrt{-g}\,\left [\L \delta_\nu^\mu \nabla_\mu \xi^\nu + 2\frac{\partial \L}{\partial g^{00}}g^{\mu 0}\delta^0_\nu \nabla_\mu \xi^\nu + \frac{\partial \L}{\partial K_{\mu \nu}}\left(\xi^\rho \nabla_\rho K_{\mu \nu} + 2K_{\nu \rho} \nabla_\mu \xi^\rho \right) \right].
\ee

Collecting the terms with $\nabla_\mu \xi^\nu$ gives the bare current $j^\mu$. The full current is $J^\mu = j^\mu - k^\mu$:

\begin{align}
J^\mu &= 2\left(\frac{\partial \L}{\partial g^{00}}g^{\mu 0}  + \frac{\partial \L}{\partial K_{\mu \nu}}K_{0 \nu} \right)  \xi^0 \\
&= \frac{2}{\dot{\bar\Phi}}\sum_{n,m\ge0} \frac{(\delta g^{00})^n\delta K_{\mu_1}^{\nu_1}\dotsi \delta K_{\mu_m}^{\nu_m}}{n!m!} \left [ (m+1)\dum{{n+1}}{{m+1}}\delta K_{\mu_{m+1}}^{\nu_{m+1}} g^{\mu0} + (n+1){\dum{{n+1}}{m}}^{\mu}_{\nu}K_0^\nu\right].
\end{align}

Note that $\nabla_i J^i = 0$ due to metric compatibility and because the functions $c$ depend on time only; thus the conservation law is simply $\nabla_0 J^0=0$.
As usual, a Noether charge can be constructed by integrating the charge density $J^0$ over a constant time slice:

\begin{multline}
Q = \frac{2}{\dot{\bar\Phi}}\sum_{n,m\ge0}\int \rd^3 \bfx \,  \sqrt h \,  \frac{(\delta g^{00})^n\delta K_{\mu_1}^{\nu_1}\dotsi \delta K_{\mu_m}^{\nu_m}}{n! \, m!} 
\\
\times\left [ (m+1)\dum{{n+1}}{{m+1}}\delta K_{\mu_{m+1}}^{\nu_{m+1}} g^{00} + (n+1){\dum{{n+1}}{m}}^{0}_{\nu}K_0^\nu\right].
\end{multline}

\section{Checking the Definition of $  \R $\label{Appendix:CheckingRInClockTime}}

In this subsection and the next, we check that the correct definition of $  \R $ (or $  \zeta $) in clock time is the same as that in cosmological time, namely
\be
\rd s^{2}=-N^{2}\rd t^{2}+a^{2}e^{2\R_{c}}\left(  e^{\gamma}\right)_{ij}\left( \rd x^{i}+N^{i}\rd t \right)\left( \rd x^{j}+N^{j}\rd t \right)\,,
\ee
in comoving gauge, $  \delta u=0 $, with $  \left( e^{\gamma} \right)_{ij} $ traceless for tensors. We call it $  \R_{c} $ because this is the value of the gauge-invariant variable $  \R $ in comoving gauge. There are two ways to proceed. One way is to just perform a large gauge transformation from cosmological time to clock time and see that $  \R $ is unchanged ($  \R $ changes only under large spatial gauge transformations). The second way is to derive the local expansion rate as in the $  \delta N $ formalism \cite{Sasaki:1995aw}.


\subsection*{Method 1: Gauge Transformations to Clock Time}

Under a coordinate transformation $  x\rightarrow x'^{\mu}=x^{\mu}+\e^{\mu}(x) $ (in $  \bp $ time), We find the gauge tranformations
\be
\Delta h_{00}&=&-2\partial_{\bp}\e_{0}+2\cH_{f}\e_{0}=2f\partial_{\bp}\e^{0}\,,\\
\Delta h_{0i}&=&-\partial_{\bp}\e_{i}-\partial_{i}\e_{0}+2\cH\e_{i}\,,\\
\Delta h_{ij}&=&-2\partial_{(i}\e_{j)}+2\cH\delta_{ij}\e_{0}\frac{a^{2}}{f^{2}}\,,\\
\frac{\Delta \delta s}{\dot{\bar s}}&=&-\e^{0}=\frac{\e_{0}}{f^{2}}\,,
\ee
for any scalar $  s $.

Consider a shift symmetric scalar field couple to gravity ($  P(X) $ as a canonical example). Clearly, for any solution $  \phi_{s}(t) $ of the full theory (including $  \phi_{s}=0 $ as a special case), there is a family of additional solutions given by
\be
\phi(x)=\phi_{sol}(x)+c\,.
\ee
What is a bit unique about shift symmetries\footnote{This is different from what happens with linear symmetries such as $  \phi\rightarrow e^{i\alpha}\phi $ with $  \phi\in \mathbb{C} $. In this case, in the presence of a background $  \bar\phi $, the transformation of perturbations $  \delta \phi=\phi-\bar\phi $ is different (putting the entire transformation into the perturbation and leaving the background unchanged, as is standard in cosmological perturbation theory)
\be
\Delta \phi=i\alpha\phi \quad \text{vs}\quad \Delta \delta\phi=i\alpha\delta \phi+i\alpha \bar\phi\,.
\ee} is that this remains the case for perturbations, with the exact same transformation:
\be
\delta \phi(x)=\delta\phi_{sol}(x)+c\,.
\ee
So we know $  \delta \phi=c $ is a solution. This is a bit confusing because
\be
\R|_{comov}=\frac{A}{2}=\frac{H}{\dot\phi}\delta \phi|_{flat}
\ee
is not a solution of the $  \zeta $ equations of motion for constant $  \delta\phi $. Let us first check that this relation is correct for the case at hand. We want to cancel $  \delta \phi $, so $  \e=\delta \phi/\dot\phi $, which is a large gauge transformation and not a small one. Under a general gauge transformation $  \e^{\mu} $ (large or small) one finds (see W 5.3.5 adn 5.3.14) 
\be
\Delta h_{ij}&=&-2\partial_{(i}\e_{j)}+2a^{2}\delta_{ij}\e_{0}\,,\label{first}\\
-\frac{\Delta\delta\rho}{\dot{\bar\rho}}&=&\Delta\delta u=-\e_{0}\,.
\ee
If we work in a gauge in which the spatial metric is diagonal (as in Newtonian, flat or comoving gauges)
\be
h_{ij}=a^{2}A\delta_{ij}\,,
\ee 
We find that for any $  \e^{\mu}=\{\e^{0}(t,x),0\} $ one gets 
\be
\Delta \zeta=\Delta \R=0\,.
\ee
$  \R $ and $  \zeta $ are not invariant only under large gauge transformations for which $  \e^{i}\neq 0 $, so that the first term on the right hand side of \refeq{first} contributes.

 
\subsection*{Method 2: Local Expansion Rate}

Consider a unit four vector $  u^{\mu}u_{\mu}=-1 $, whose spatial part is first order in perturbation. The normalization implies 
\be\label{obs}
u^{\mu}=\frac{1}{f}\{1+\frac{h_{00}}{2f^{2}},v^{i}\}\,.
\ee
We want to compute the geodesic convergence $ \theta\equiv \nabla_{\mu}u^{\mu} $. To zeroth order this gives
\be
\theta_{0}=\frac{3\cH}{f}=3H \quad \text{(0-th order)}\,,
\ee
which agree with the expectation that $  \theta/3 $ is the locally measured expansion rate. To first order, We find
\be
\theta_{1}=\frac{1}{f}\left[  \frac{3}{2}\cH \frac{h_{00}}{f^{2}}+\frac{1}{2a^{2}}\left(  h_{ii}'-2\cH h_{ii}\right)+\partial_{i}v^{i}\right]\,,
\ee
which agrees with Eq (4.8) of \cite{Dai:2015rda}, after setting $  f=a $ and accounting for the different notation $  a^{2}h_{\mu\nu}^{there}=h_{\mu\nu}^{here} $. It is perhaps remarkable that $  f $ appears without derivatives in the local expansion rate. To get the total local expansion, we want to integrate the local expansion rate in proper time
\be
\cN=\int \rd t_{p} \frac{\theta}{3}\,.
\ee
The proper time of the observer \eqref{obs}, assuming it is at rest in comoving coordinates, $  v^{i}=0 $ is 
\be
\rd t_{p}=\sqrt{-\rd x^{\mu}g_{\mu\nu}\rd x^{\nu}}=\sqrt{-g_{00}}\rd\bp=f\rd\bp\left(  1-\frac{h_{00}}{2f^{2}}\right)\,.
\ee
This had to be the case since we know that FLRW time represent proper time of comoving observers. Finally, the perturbations that are conserved on superHubble scales and that we use for late time cosmology are
\be
\cN-\cN_{0}&=&\int f \rd\bp \frac{1}{3f}\left[  \frac{3}{2}\cH \frac{h_{00}}{f^{2}} (1-1)+\frac{1}{2a^{2}}\left(  h_{ii}'-2\cH h_{ii}\right)\right]\\
&=&\int \rd\bp \frac{1}{6a^{2}}\left(  h_{ii}'-2\cH h_{ii}\right)\,.
\ee
If we parameterize the perturbations to the trace of spatial metric by
\be
g_{ij}=a^{2}\delta_{ij}\left(  1+A \right)+\dots \,,
\ee
we find 
\be
\cN(\phi)-\cN_{0}&=&\frac{1}{2}\int^{\phi} \rd\phi' \partial_{\phi}A(\phi')=\frac{1}{2}\left[ A(\phi)-A(\phi_{i}) \right]\,,
\ee
which is the same formula as in standard FLRW coordinates, upon the identification $  \R=A/2 $.


\section{Clock Time EFT\label{App:ClockTimeEFT}}

In this Appendix, we expand upon the EFT of shift-symmetric cosmologies using the clock time coordinates introduced in Sec.~\ref{Sec:SSandClockTime}.  

The leading EFT terms in clock time are of the standard form:
\begin{align}
 S_{\delta g}&=\int\rd\phi\rd^{3}x\, \sqrt{-g}\sum_{n}\frac{d_{n}(\phi)}{n!}\left (\delta g^{00}\right )^{n}
\end{align}
where the $\delta g^{00}$'s now correspond to fluctuations about the metric \eqref{ClockTimeFRW}
\begin{align}
\rd s^{2}&=-\frac{\rd \phi^{2}}{f(\phi)^{2}}+a(\phi)^{2}\rd \vec{x}^{2}\ .
\end{align}
We will not consider $\sim \left (\delta K\right )^{n}$ and higher derivative operators here.  These coordinates are chosen such that the scalar field solution is $\bar{\Phi}(x^{\mu})=\phi$ exactly, thus the diffeomorphism that represents a residual symmetry of the unitary gauge action is simply:
\begin{align}
x^{\mu}\to x'^{\mu}=x^{\mu}+c\delta^{\mu}_{0}\ ,
\end{align}
for constant $c$.  Following the procedure in Sec.~\ref{Sec:Sofg00Relations}, this diffeomorphism can be shown to change the action by
\begin{align}
\Delta S_{\delta g}&=c\int\rd\phi\rd^{3}x\, \sqrt{-g}\sum_{n}\frac{\left (d'_{n}(\phi)+2f'(\phi)f(\phi)d_{n+1}(\phi)\right )}{n!}\left (\delta g^{00}\right )^{n}
\end{align}
and hence a shift symmetry corresponds to the following clock time recursion relations:
\begin{align}
d'_{n}(\phi)&=-2f'(\phi)f(\phi)d_{n+1}(\phi)\ .\label{ClockTimeRecursionRelations}
\end{align}

Next, we discuss the Goldstone theory in clock time.  We will show that $\varphi\to \varphi+c$ is a symmetry in these coordinates. As discussed in Sec.~\ref{subsubsec:ClockTimeGoldstones}, the action is given by \eqref{DeltaGActionClockTime2}:
\begin{align}
S_{\pi}&=\int\rd^{4}x\sqrt{-g}\, \sum_{n}\frac{ d_{n}(\phi+\varphi)}{n!}\left (f^{2}(\phi)\left (-2\partial_{\phi}\varphi-(\partial_{\phi}\varphi)^{2}\right )+\frac{(\partial_{i}\varphi)^{2}}{a^{2}(\phi)}+f^{2}(\phi+\varphi)-f^{2}(\phi)\right )^{n}\nn
&\equiv\int\rd^{4}x\sqrt{-g}\, \sum_{n}\frac{ d_{n}(\phi+\varphi)}{n!}\left (\delta g^{00}(\varphi))\right )^{n} \ ,\label{DeltaGActionClockTime3}
\end{align}
where we introduced useful shorthand for the St\"uckelberged form of $\delta g^{00}$.  If we now replace $\varphi\to \varphi+c$, the above changes by
\begin{align}
\Delta S_{\pi}&=c\int\rd^{4}x\sqrt{-g}\, \sum_{n}\frac{ d_{n}'(\phi+\varphi)}{n!}\left (\delta g^{00}(\varphi))\right )^{n}+\frac{ d_{n}(\phi+\varphi)}{(n-1)!}\left (\delta g^{00}(\varphi))\right )^{n-1}\left (2f'(\phi+\varphi)f(\phi+\varphi)\right )\nn
&=c\int\rd^{4}x\sqrt{-g}\, \sum_{n}\frac{\Big (d_{n}'(\phi+\varphi)+2f'(\phi+\varphi)f(\phi+\varphi)d_{n+1}(\phi+\varphi)\Big )}{n!}\left (\delta g^{00}(\varphi))\right )^{n}\ ,
\end{align}
which vanishes for shift symmetric theories due to \eqref{ClockTimeRecursionRelations}. In such cases, it is therefore possible to integrate the action \eqref{DeltaGActionClockTime3} by parts such that $\varphi$ always appears with at least one derivative acting upon it.


\section{No Slow-Roll in $P(X)$ }\label{App:PofXSection}

In this Appendix, we use the $P(X)$ Lagrangian to directly re-derive the relation \eqref{beomsst} which demonstrates the absence of slow-roll solutions in this class of models.

We start by briefly reviewing the $P(X)$ equations of motion. The Lagrangian is written as\footnote{For dimensional reasons, another scale $\Lambda$ is required to define a $P(X)$ theory.  Here, we will set this scale to unity for clarity of presentation.}
\begin{align}
\mathcal{L}=P(X)\ , \quad X\equiv -(\partial\Phi)^{2}\ \label{PXLagrangian}
\end{align}
and the associated stress tensor is given by:
\begin{align}
T_{\mu\nu}=-\frac{2}{\sqrt{-g}}\frac{\delta S_{M}}{\delta g^{\mu\nu}}=g_{\mu\nu}P(X)+2\nabla_{\mu}\Phi\nabla_{\nu}\Phi P'(X)\ .
\end{align}
Introducing the 4-velocity $u_{\mu}=\nabla_{\mu}\Phi/\sqrt{-(\partial\Phi)^{2}}$, the above can be written in perfect fluid form:
\begin{align}
T_{\mu\nu}&=(\rho+p)u_{\mu}u_{\nu}+pg_{\mu\nu}\ , \quad p=P \ , \quad \rho=-P+2XP'(X)\ .
\end{align}
The $\Phi$ equation of motion is
\begin{align}
0&=2\square\Phi P'(X)-4\nabla^{\mu}\Phi\nabla_{\mu}\nabla_{\nu}\Phi\nabla^{\nu}\Phi P''(X)\ \label{PXPhiEOM}
\end{align}
and evaluating on a $\Phi=\Phi(t)$ ansatz, this can be conveniently re-written as
\begin{align}
0&=\left (6H X+\dot{X}\right )P'(X)+2X \dot{ X}P''(X)\ .\label{PXPhiEOMPhiOfT}
\end{align}Finally, the two independent gravitational equations are
\begin{align}
0&=H^{2}-\frac{1}{2}H^{2}\varepsilon+\frac{1}{\Mpl^{2}}P(X)-\frac{X}{6\Mpl^{2}}P'(X)\,,\nn
0&=\frac{3}{2}H^{2}-H^{2}\varepsilon+\frac{1}{2\Mpl^{2}}P(X)\ .\label{PXGravEOM}
\end{align}

Our goal is to re-write \eqref{PXPhiEOMPhiOfT} in terms of $H$, $\varepsilon, \eta$ and $ c_{s}^{2}$.
First, the gravitational EOM can be used to solve for $P(X)$ and its derivative:
\begin{align}
P(X)=\Mpl^{2}H^{2}(2\varepsilon-3)\ , \quad XP'(X)=\Mpl^{2}H^{2}\varepsilon\ .\label{Plug1}
\end{align}
The speed of sound is calculated via $c_{s}^{2}=\partial_{X}p/\partial_{X}\rho$, which can be used to replace $P''(X)$:
\begin{align}
c_{s}^{2}=\frac{\varepsilon}{\varepsilon+\frac{2X^{2}P''(X)}{\Mpl^{2}H^{2}}}
\implies X^{2}P''(X)=\frac{\varepsilon}{2} \Mpl^{2}H^{2}\left (1-\frac{1}{c_{s}^{2}}\right )\ .\label{Plug2}
\end{align}
The final ingredient is an expression for $\dot{ X}$.  This comes from taking a derivative of either gravitational equation of motion, leading to
\begin{align}
\dot{X}&=2 H X\left (3+\eta-2\varepsilon\right )\ .\label{Plug3}
\end{align}

Plugging \eqref{Plug1}, \eqref{Plug2} and \eqref{Plug3} into \eqref{PXPhiEOMPhiOfT}, we are left with
\begin{align}
0&=\frac{\Mpl^{2}H^{3}}{c_{s}^{2}\sqrt{X}}\varepsilon\left (3+3c_{s}^{2}-2\varepsilon+\eta\right )\ ,\label{NoSlowRollConditionAppendix}
\end{align}
which is equivalent to \eqref{beomsst}.

\bibliographystyle{utphys}
\bibliography{ShiftSymmetricEFTOfInflation}

\providecommand{\href}[2]{#2}\begingroup\raggedright\begin{thebibliography}{10}

\bibitem{Baumann:2014nda}
D.~Baumann and L.~McAllister, {\em {Inflation and String Theory}}.
\newblock Cambridge University Press, 2015.
\newblock \href{http://arxiv.org/abs/1404.2601}{{\tt arXiv:1404.2601
  [hep-th]}}.
\newblock
\url{http://inspirehep.net/record/1289899/files/arXiv:1404.2601.pdf}.
\newblock

\bibitem{ArkaniHamed:2006dz}
N.~Arkani-Hamed, L.~Motl, A.~Nicolis, and C.~Vafa, ``{The String landscape,
  black holes and gravity as the weakest force},''
  \href{http://dx.doi.org/10.1088/1126-6708/2007/06/060}{{\em JHEP} {\bf 0706}
  (2007)  060},
\href{http://arxiv.org/abs/hep-th/0601001}{{\tt arXiv:hep-th/0601001
  [hep-th]}}.

\bibitem{Kallosh:1995hi}
R.~Kallosh, A.~D. Linde, D.~A. Linde, and L.~Susskind, ``{Gravity and global
  symmetries},'' \href{http://dx.doi.org/10.1103/PhysRevD.52.912}{{\em Phys.
  Rev.} {\bf D52} (1995)  912--935},
\href{http://arxiv.org/abs/hep-th/9502069}{{\tt arXiv:hep-th/9502069
  [hep-th]}}.

\bibitem{Banks:2010zn}
T.~Banks and N.~Seiberg, ``{Symmetries and Strings in Field Theory and
  Gravity},'' \href{http://dx.doi.org/10.1103/PhysRevD.83.084019}{{\em Phys.
  Rev.} {\bf D83} (2011)  084019},
\href{http://arxiv.org/abs/1011.5120}{{\tt arXiv:1011.5120 [hep-th]}}.

\bibitem{Vafa:2005ui}
C.~Vafa, ``{The String landscape and the swampland},''
\href{http://arxiv.org/abs/hep-th/0509212}{{\tt arXiv:hep-th/0509212
  [hep-th]}}.

\bibitem{Berg:2009tg}
M.~Berg, E.~Pajer, and S.~Sjors, ``{Dante's Inferno},''
  \href{http://dx.doi.org/10.1103/PhysRevD.81.103535}{{\em Phys. Rev.} {\bf
  D81} (2010)  103535},
\href{http://arxiv.org/abs/0912.1341}{{\tt arXiv:0912.1341 [hep-th]}}.

\bibitem{Cheung:2007st}
C.~Cheung, P.~Creminelli, A.~L. Fitzpatrick, J.~Kaplan, and L.~Senatore, ``{The
  Effective Field Theory of Inflation},''
  \href{http://dx.doi.org/10.1088/1126-6708/2008/03/014}{{\em JHEP} {\bf 0803}
  (2008)  014},
\href{http://arxiv.org/abs/0709.0293}{{\tt arXiv:0709.0293 [hep-th]}}.

\bibitem{InPreparation}
B.~Finelli, G.~Goon, E.~Pajer, and L.~Santoni, ``{In Preparation},''.

\bibitem{Berezhiani:2015bqa}
L.~Berezhiani and J.~Khoury, ``{Theory of dark matter superfluidity},''
  \href{http://dx.doi.org/10.1103/PhysRevD.92.103510}{{\em Phys. Rev.} {\bf
  D92} (2015)  103510},
\href{http://arxiv.org/abs/1507.01019}{{\tt arXiv:1507.01019 [astro-ph.CO]}}.

\bibitem{Celoria:2017bbh}
M.~Celoria, D.~Comelli, and L.~Pilo, ``{Fluids, Superfluids and Supersolids:
  Dynamics and Cosmology of Self Gravitating Media},''
  \href{http://dx.doi.org/10.1088/1475-7516/2017/09/036}{{\em JCAP} {\bf 1709}
  (2017) no.~09, 036},
\href{http://arxiv.org/abs/1704.00322}{{\tt arXiv:1704.00322 [gr-qc]}}.

\bibitem{Celoria:2017xos}
M.~Celoria, D.~Comelli, and L.~Pilo, ``{Intrinsic Entropy Perturbations from
  the Dark Sector},''
\href{http://arxiv.org/abs/1711.01961}{{\tt arXiv:1711.01961 [gr-qc]}}.

\bibitem{Nicolis:2011pv}
A.~Nicolis and F.~Piazza, ``{Spontaneous Symmetry Probing},''
  \href{http://dx.doi.org/10.1007/JHEP06(2012)025}{{\em JHEP} {\bf 06} (2012)
  025},
\href{http://arxiv.org/abs/1112.5174}{{\tt arXiv:1112.5174 [hep-th]}}.

\bibitem{Finelli:2017fml}
B.~Finelli, G.~Goon, E.~Pajer, and L.~Santoni, ``{Soft Theorems For
  Shift-Symmetric Cosmologies},''
\href{http://arxiv.org/abs/1711.03737}{{\tt arXiv:1711.03737 [hep-th]}}.

\bibitem{Mooij:2015yka}
S.~Mooij and G.~A. Palma, ``{Consistently violating the non-Gaussian
  consistency relation},''
  \href{http://dx.doi.org/10.1088/1475-7516/2015/11/025}{{\em JCAP} {\bf 1511}
  (2015) no.~11, 025},
\href{http://arxiv.org/abs/1502.03458}{{\tt arXiv:1502.03458 [astro-ph.CO]}}.

\bibitem{Bravo:2017wyw}
R.~Bravo, S.~Mooij, G.~A. Palma, and B.~Pradenas, ``{A generalized non-Gaussian
  consistency relation for single field inflation},''
\href{http://arxiv.org/abs/1711.02680}{{\tt arXiv:1711.02680 [astro-ph.CO]}}.

\bibitem{Nicolis:2015sra}
A.~Nicolis, R.~Penco, F.~Piazza, and R.~Rattazzi, ``{Zoology of condensed
  matter: Framids, ordinary stuff, extra-ordinary stuff},''
  \href{http://dx.doi.org/10.1007/JHEP06(2015)155}{{\em JHEP} {\bf 06} (2015)
  155},
\href{http://arxiv.org/abs/1501.03845}{{\tt arXiv:1501.03845 [hep-th]}}.

\bibitem{Nicolis:2011cs}
A.~Nicolis, ``{Low-energy effective field theory for finite-temperature
  relativistic superfluids},''
\href{http://arxiv.org/abs/1108.2513}{{\tt arXiv:1108.2513 [hep-th]}}.

\bibitem{Son:2002zn}
D.~T. Son, ``{Low-energy quantum effective action for relativistic
  superfluids},''
\href{http://arxiv.org/abs/hep-ph/0204199}{{\tt arXiv:hep-ph/0204199
  [hep-ph]}}.

\bibitem{Dubovsky:2005xd}
S.~Dubovsky, T.~Gregoire, A.~Nicolis, and R.~Rattazzi, ``{Null energy condition
  and superluminal propagation},''
  \href{http://dx.doi.org/10.1088/1126-6708/2006/03/025}{{\em JHEP} {\bf 03}
  (2006)  025},
\href{http://arxiv.org/abs/hep-th/0512260}{{\tt arXiv:hep-th/0512260
  [hep-th]}}.

\bibitem{Carter:1999zw}
B.~Carter, ``{Relativistic dynamics of vortex defects in superfluids},'' in
  {\em {Topological defects and the nonequilibrium dynamics of symmetry
  breaking phase transitions. Proceedings, NATO Advanced Study Institute, ESF
  Network Workshop and Winter School, Les Houches, France, February 16-26,
  1999}}, pp.~267--301.
\newblock 1999.
\newblock
\href{http://arxiv.org/abs/gr-qc/9907039}{{\tt arXiv:gr-qc/9907039 [gr-qc]}}.
\newblock

\bibitem{khalatnikov1982relativistic}
I.~Khalatnikov and V.~Lebedev, ``Relativistic hydrodynamics of a superfluid
  liquid,'' {\em Physics Letters A} {\bf 91} (1982) no.~2, 70--72.

\bibitem{Nicolis:2013lma}
A.~Nicolis, R.~Penco, and R.~A. Rosen, ``{Relativistic Fluids, Superfluids,
  Solids and Supersolids from a Coset Construction},''
  \href{http://dx.doi.org/10.1103/PhysRevD.89.045002}{{\em Phys. Rev.} {\bf
  D89} (2014) no.~4, 045002},
\href{http://arxiv.org/abs/1307.0517}{{\tt arXiv:1307.0517 [hep-th]}}.

\bibitem{Ogievetsky}
V.~I. Ogievetsky, ``{ Nonlinear Realizations of Internal and Space-time
  Symmetries },'' {\em Proc. of. X-th Winter. School of Theoretical Physics in
  Karpacz, Vol. 1, Wroclaw 227 (1974)}  .

\bibitem{Pujolas:2011he}
O.~Pujolas, I.~Sawicki, and A.~Vikman, ``{The Imperfect Fluid behind Kinetic
  Gravity Braiding},'' \href{http://dx.doi.org/10.1007/JHEP11(2011)156}{{\em
  JHEP} {\bf 11} (2011)  156},
\href{http://arxiv.org/abs/1103.5360}{{\tt arXiv:1103.5360 [hep-th]}}.

\bibitem{Deffayet:2010qz}
C.~Deffayet, O.~Pujolas, I.~Sawicki, and A.~Vikman, ``{Imperfect Dark Energy
  from Kinetic Gravity Braiding},''
  \href{http://dx.doi.org/10.1088/1475-7516/2010/10/026}{{\em JCAP} {\bf 1010}
  (2010)  026},
\href{http://arxiv.org/abs/1008.0048}{{\tt arXiv:1008.0048 [hep-th]}}.

\bibitem{Creminelli:2006xe}
P.~Creminelli, M.~A. Luty, A.~Nicolis, and L.~Senatore, ``{Starting the
  Universe: Stable Violation of the Null Energy Condition and Non-standard
  Cosmologies},'' \href{http://dx.doi.org/10.1088/1126-6708/2006/12/080}{{\em
  JHEP} {\bf 12} (2006)  080},
\href{http://arxiv.org/abs/hep-th/0606090}{{\tt arXiv:hep-th/0606090
  [hep-th]}}.

\bibitem{Pirtskhalava:2015nla}
D.~Pirtskhalava, L.~Santoni, E.~Trincherini, and F.~Vernizzi, ``{Weakly Broken
  Galileon Symmetry},''
  \href{http://dx.doi.org/10.1088/1475-7516/2015/09/007}{{\em JCAP} {\bf 1509}
  (2015) no.~09, 007},
\href{http://arxiv.org/abs/1505.00007}{{\tt arXiv:1505.00007 [hep-th]}}.

\bibitem{Goon:2016ihr}
G.~Goon, K.~Hinterbichler, A.~Joyce, and M.~Trodden, ``{Aspects of Galileon
  Non-Renormalization},'' \href{http://dx.doi.org/10.1007/JHEP11(2016)100}{{\em
  JHEP} {\bf 11} (2016)  100},
\href{http://arxiv.org/abs/1606.02295}{{\tt arXiv:1606.02295 [hep-th]}}.

\bibitem{Pirtskhalava:2015zwa}
D.~Pirtskhalava, L.~Santoni, E.~Trincherini, and F.~Vernizzi, ``{Large
  Non-Gaussianity in Slow-Roll Inflation},''
  \href{http://dx.doi.org/10.1007/JHEP04(2016)117}{{\em JHEP} {\bf 04} (2016)
  117},
\href{http://arxiv.org/abs/1506.06750}{{\tt arXiv:1506.06750 [hep-th]}}.

\bibitem{Pirtskhalava:2015ebk}
D.~Pirtskhalava, L.~Santoni, and E.~Trincherini, ``{Constraints on Single-Field
  Inflation},'' \href{http://dx.doi.org/10.1088/1475-7516/2016/06/051}{{\em
  JCAP} {\bf 1606} (2016) no.~06, 051},
\href{http://arxiv.org/abs/1511.01817}{{\tt arXiv:1511.01817 [hep-th]}}.

\bibitem{Horndeski:1974wa}
G.~W. Horndeski, ``{Second-order scalar-tensor field equations in a
  four-dimensional space},''
\href{http://dx.doi.org/10.1007/BF01807638}{{\em Int. J. Theor. Phys.} {\bf 10}
  (1974)  363--384}.

\bibitem{Deffayet:2011gz}
C.~Deffayet, X.~Gao, D.~A. Steer, and G.~Zahariade, ``{From k-essence to
  generalised Galileons},''
  \href{http://dx.doi.org/10.1103/PhysRevD.84.064039}{{\em Phys. Rev.} {\bf
  D84} (2011)  064039},
\href{http://arxiv.org/abs/1103.3260}{{\tt arXiv:1103.3260 [hep-th]}}.

\bibitem{Langlois:2015cwa}
D.~Langlois and K.~Noui, ``{Degenerate higher derivative theories beyond
  Horndeski: evading the Ostrogradski instability},''
  \href{http://dx.doi.org/10.1088/1475-7516/2016/02/034}{{\em JCAP} {\bf 1602}
  (2016) no.~02, 034},
\href{http://arxiv.org/abs/1510.06930}{{\tt arXiv:1510.06930 [gr-qc]}}.

\bibitem{Kobayashi:2010cm}
T.~Kobayashi, M.~Yamaguchi, and J.~Yokoyama, ``{G-inflation: Inflation driven
  by the Galileon field},''
  \href{http://dx.doi.org/10.1103/PhysRevLett.105.231302}{{\em Phys. Rev.
  Lett.} {\bf 105} (2010)  231302},
\href{http://arxiv.org/abs/1008.0603}{{\tt arXiv:1008.0603 [hep-th]}}.

\bibitem{Burrage:2010cu}
C.~Burrage, C.~de~Rham, D.~Seery, and A.~J. Tolley, ``{Galileon inflation},''
  \href{http://dx.doi.org/10.1088/1475-7516/2011/01/014}{{\em JCAP} {\bf 1101}
  (2011)  014},
\href{http://arxiv.org/abs/1009.2497}{{\tt arXiv:1009.2497 [hep-th]}}.

\bibitem{Gao:2011qe}
X.~Gao and D.~A. Steer, ``{Inflation and primordial non-Gaussianities of
  'generalized Galileons'},''
  \href{http://dx.doi.org/10.1088/1475-7516/2011/12/019}{{\em JCAP} {\bf 1112}
  (2011)  019},
\href{http://arxiv.org/abs/1107.2642}{{\tt arXiv:1107.2642 [astro-ph.CO]}}.

\bibitem{DeFelice:2013ar}
A.~De~Felice and S.~Tsujikawa, ``{Shapes of primordial non-Gaussianities in the
  Horndeski's most general scalar-tensor theories},''
  \href{http://dx.doi.org/10.1088/1475-7516/2013/03/030}{{\em JCAP} {\bf 1303}
  (2013)  030},
\href{http://arxiv.org/abs/1301.5721}{{\tt arXiv:1301.5721 [hep-th]}}.

\bibitem{Behbahani:2011it}
S.~R. Behbahani, A.~Dymarsky, M.~Mirbabayi, and L.~Senatore, ``{(Small)
  Resonant non-Gaussianities: Signatures of a Discrete Shift Symmetry in the
  Effective Field Theory of Inflation},''
  \href{http://dx.doi.org/10.1088/1475-7516/2012/12/036}{{\em JCAP} {\bf 1212}
  (2012)  036},
\href{http://arxiv.org/abs/1111.3373}{{\tt arXiv:1111.3373 [hep-th]}}.

\bibitem{ArkaniHamed:2003uz}
N.~Arkani-Hamed, P.~Creminelli, S.~Mukohyama, and M.~Zaldarriaga, ``{Ghost
  inflation},'' \href{http://dx.doi.org/10.1088/1475-7516/2004/04/001}{{\em
  JCAP} {\bf 0404} (2004)  001},
\href{http://arxiv.org/abs/hep-th/0312100}{{\tt arXiv:hep-th/0312100
  [hep-th]}}.

\bibitem{Kinney:2005vj}
W.~H. Kinney, ``{Horizon crossing and inflation with large eta},''
  \href{http://dx.doi.org/10.1103/PhysRevD.72.023515}{{\em Phys. Rev.} {\bf
  D72} (2005)  023515},
\href{http://arxiv.org/abs/gr-qc/0503017}{{\tt arXiv:gr-qc/0503017 [gr-qc]}}.

\bibitem{Bravo:2017gct}
R.~Bravo, S.~Mooij, G.~A. Palma, and B.~Pradenas, ``{Vanishing of local
  non-Gaussianity in canonical single field inflation},''
\href{http://arxiv.org/abs/1711.05290}{{\tt arXiv:1711.05290 [astro-ph.CO]}}.

\bibitem{Cai:2017bxr}
Y.-F. Cai, X.~Chen, M.~H. Namjoo, M.~Sasaki, D.-G. Wang, and Z.~Wang,
  ``{Revisiting non-Gaussianity from non-attractor inflation models},''
\href{http://arxiv.org/abs/1712.09998}{{\tt arXiv:1712.09998 [astro-ph.CO]}}.

\bibitem{ArkaniHamed:2003uy}
N.~Arkani-Hamed, H.-C. Cheng, M.~A. Luty, and S.~Mukohyama, ``{Ghost
  condensation and a consistent infrared modification of gravity},''
  \href{http://dx.doi.org/10.1088/1126-6708/2004/05/074}{{\em JHEP} {\bf 05}
  (2004)  074},
\href{http://arxiv.org/abs/hep-th/0312099}{{\tt arXiv:hep-th/0312099
  [hep-th]}}.

\bibitem{Weinberg:2003sw}
S.~Weinberg, ``{Adiabatic modes in cosmology},''
  \href{http://dx.doi.org/10.1103/PhysRevD.67.123504}{{\em Phys. Rev.} {\bf
  D67} (2003)  123504},
\href{http://arxiv.org/abs/astro-ph/0302326}{{\tt arXiv:astro-ph/0302326
  [astro-ph]}}.

\bibitem{Pajer:2017hmb}
E.~Pajer and S.~Jazayeri, ``{Systematics of Adiabatic Modes: Flat Universes},''
\href{http://arxiv.org/abs/1710.02177}{{\tt arXiv:1710.02177 [astro-ph.CO]}}.

\bibitem{Hinterbichler:2013dpa}
K.~Hinterbichler, L.~Hui, and J.~Khoury, ``{An Infinite Set of Ward Identities
  for Adiabatic Modes in Cosmology},''
  \href{http://dx.doi.org/10.1088/1475-7516/2014/01/039}{{\em JCAP} {\bf 1401}
  (2014)  039},
\href{http://arxiv.org/abs/1304.5527}{{\tt arXiv:1304.5527 [hep-th]}}.

\bibitem{Senatore:2009gt}
L.~Senatore, K.~M. Smith, and M.~Zaldarriaga, ``{Non-Gaussianities in Single
  Field Inflation and their Optimal Limits from the WMAP 5-year Data},''
  \href{http://dx.doi.org/10.1088/1475-7516/2010/01/028}{{\em JCAP} {\bf 1001}
  (2010)  028},
\href{http://arxiv.org/abs/0905.3746}{{\tt arXiv:0905.3746 [astro-ph.CO]}}.

\bibitem{Ade:2015ava}
{\bf Planck} Collaboration, P.~A.~R. Ade {\em et al.}, ``{Planck 2015 results.
  XVII. Constraints on primordial non-Gaussianity},''
  \href{http://dx.doi.org/10.1051/0004-6361/201525836}{{\em Astron. Astrophys.}
  {\bf 594} (2016)  A17},
\href{http://arxiv.org/abs/1502.01592}{{\tt arXiv:1502.01592 [astro-ph.CO]}}.

\bibitem{Baumann:2011su}
D.~Baumann and D.~Green, ``{Equilateral Non-Gaussianity and New Physics on the
  Horizon},'' \href{http://dx.doi.org/10.1088/1475-7516/2011/09/014}{{\em JCAP}
  {\bf 1109} (2011)  014},
\href{http://arxiv.org/abs/1102.5343}{{\tt arXiv:1102.5343 [hep-th]}}.

\bibitem{Bartolo:2015qvr}
N.~Bartolo, D.~Cannone, A.~Ricciardone, and G.~Tasinato, ``{Distinctive
  signatures of space-time diffeomorphism breaking in EFT of inflation},''
  \href{http://dx.doi.org/10.1088/1475-7516/2016/03/044}{{\em JCAP} {\bf 1603}
  (2016) no.~03, 044},
\href{http://arxiv.org/abs/1511.07414}{{\tt arXiv:1511.07414 [astro-ph.CO]}}.

\bibitem{Lin:2015cqa}
C.~Lin and L.~Z. Labun, ``{Effective Field Theory of Broken Spatial
  Diffeomorphisms},'' \href{http://dx.doi.org/10.1007/JHEP03(2016)128}{{\em
  JHEP} {\bf 03} (2016)  128},
\href{http://arxiv.org/abs/1501.07160}{{\tt arXiv:1501.07160 [hep-th]}}.

\bibitem{Baumann:2011ws}
D.~Baumann and D.~Green, ``{A Field Range Bound for General Single-Field
  Inflation},'' \href{http://dx.doi.org/10.1088/1475-7516/2012/05/017}{{\em
  JCAP} {\bf 1205} (2012)  017},
\href{http://arxiv.org/abs/1111.3040}{{\tt arXiv:1111.3040 [hep-th]}}.

\bibitem{Sasaki:1995aw}
M.~Sasaki and E.~D. Stewart, ``{A General analytic formula for the spectral
  index of the density perturbations produced during inflation},''
  \href{http://dx.doi.org/10.1143/PTP.95.71}{{\em Prog. Theor. Phys.} {\bf 95}
  (1996)  71--78},
\href{http://arxiv.org/abs/astro-ph/9507001}{{\tt arXiv:astro-ph/9507001
  [astro-ph]}}.

\bibitem{Dai:2015rda}
L.~Dai, E.~Pajer, and F.~Schmidt, ``{Conformal Fermi Coordinates},''
  \href{http://dx.doi.org/10.1088/1475-7516/2015/11/043}{{\em JCAP} {\bf 1511}
  (2015) no.~11, 043},
\href{http://arxiv.org/abs/1502.02011}{{\tt arXiv:1502.02011 [gr-qc]}}.

\end{thebibliography}\endgroup

\end{document}